\documentclass[10pt]{article}
\usepackage[a4paper,nohead]{geometry}
\usepackage[utf8x]{inputenc}
\usepackage[T1]{fontenc}
\usepackage{graphicx}
\usepackage{amsmath}
\usepackage[sort&compress,numbers]{natbib}
\usepackage{amsfonts}
\usepackage{amssymb}
\usepackage{amsmath}
\usepackage{latexsym}
\usepackage{color}
\usepackage{ntheorem}
\usepackage{braket}
\usepackage[colorlinks=true]{hyperref}
\usepackage{authblk}
\usepackage{tensor}

\newcounter{mnotecount}[section]

\renewcommand{\themnotecount}{\thesection.\arabic{mnotecount}}
\newcommand{\mnote}[1]
{\protect{\stepcounter{mnotecount}}$^{\mbox{\footnotesize
$
\bullet$\themnotecount}}$ \marginpar{
\raggedright\tiny\em
$\!\!\!\!\!\!\,\bullet$\themnotecount: #1} }

\newtheorem{theorem}{\sc  Theorem\rm}[section]

\newtheorem{lemma}[theorem]{\sc Lemma\rm}

\newtheorem{remark}[theorem]{\sc Remark\rm}

\newcommand{\ol}[1]{\overline{#1}{}}
\newcommand{\ul}[1]{\underline{#1}{}}

\newcommand{\jlcax}[1]{}
\newcommand{\eean}{\nonumber\end{eqnarray}}

\newcommand{\kk}[1]{}

\newcommand{\beq}{\begin{equation}}

\newcommand{\FS}       
                  {F}

\newcommand{\HS} 
       {H_{\mbox{\scriptsize volume}}}

{\ptc{this should be removed in the oberwolfach version}}%

\newcommand{\eeal}[1]{\label{#1}\end{eqnarray}}
\newcommand{\bed}{\begin{deqarr}}
\newcommand{\eed}{\end{deqarr}}
\newcommand{\bedl}[1]{\begin{deqarr}\label{#1}}
\newcommand{\eedl}[2]{\arrlabel{#1}\label{#2}\end{deqarr}}

\newcommand{\bel}[1]{\begin{equation}\label{#1}}
\newcommand{\bea}{\begin{eqnarray}}
\newcommand{\bean}{\begin{eqnarray}\nonumber}
\newcommand{\beal}[1]{\begin{eqnarray}\label{#1}}
\newcommand{\eea}{\end{eqnarray}}

\newcommand{\be}{\begin{equation}}
\newcommand{\eeq}{\end{equation}}
\newcommand{\ee}{\end{equation}}
\newcommand{\beqa}{\begin{eqnarray}}
\newcommand{\eeqa}{\end{eqnarray}}
\newcommand{\beqan}{\begin{eqnarray*}}
\newcommand{\eeqan}{\end{eqnarray*}}
\newcommand{\ba}{\begin{array}}
\newcommand{\ea}{\end{array}}

\newcommand{\mcM}{{\mycal M}}
\newcommand{\mcD}{{\mycal D}}

\newcommand{\scri}{{\mycal I}}%

\newcommand{\warn}[1]
{\protect{\stepcounter{mnotecount}}$^{\mbox{\footnotesize
$
\bullet$\themnotecount}}$ \marginpar{
\raggedright\tiny\em
$\!\!\!\!\!\!\,\bullet$\themnotecount: {\bf Warning:} #1} }

\newcommand{\eq}[1]{(\ref{#1})}

\newcommand{\ptc}[1]{\mnote{{\bf ptc:}#1}}

\newcommand{\beqar}{\begin{deqarr}}
\newcommand{\eeqar}{\end{deqarr}}

\newcommand{\beaa}{\begin{eqnarray*}}
\newcommand{\eeaa}{\end{eqnarray*}}

\DeclareFontFamily{OT1}{rsfs}{}
\DeclareFontShape{OT1}{rsfs}{m}{n}{ <-7> rsfs5 <7-10> rsfs7 <10-> rsfs10}{}
\DeclareMathAlphabet{\mycal}{OT1}{rsfs}{m}{n}

{\catcode `\@=11 \global\let\AddToReset=\@addtoreset}
\AddToReset{equation}{section}

{\catcode `\@=11 \global\let\AddToReset=\@addtoreset}
\AddToReset{figure}{section}

{\catcode `\@=11 \global\let\AddToReset=\@addtoreset}
\AddToReset{table}{section}

\newcommand{\rnabla}{\not\hspace{-.2em}\nabla}

\begin{document}

\title{On the choice of a conformal Gauss gauge near  the cylinder representing  spatial infinity\protect
}
\author{
Tim-Torben Paetz
\thanks{E-mail:  Tim-Torben.Paetz@univie.ac.at}  \vspace{0.5em}\\  \textit{Gravitational Physics, University of Vienna}  \\ \textit{Boltzmanngasse 5, 1090 Vienna, Austria }
}
\maketitle

\maketitle


\begin{abstract}
A convenient approach to analyze spatial infinity is to use a cylinder representation $I$ and impose a gauge based on a congruence
of conformal geodesics.
This so-called conformal Gauss gauge comes along with the freedom to specify initial data for the conformal geodesics.
Such a gauge has been constructed from an ordinary Cauchy surface and from past null infinity $\scri^-$, respectively.
The purpose of this note is to compare these gauges near the critical set $I^-$, where $I$ ``touches'' $\scri^-$, as it turns out that they are related in a somewhat  unexpected
intricate way.
\end{abstract}

\tableofcontents

\section{Introduction}

According to Penrose \cite{p1, p2} a spacetime $(\widetilde \mcM, \widetilde g)$ should be regarded as \emph{asymptotically flat} if, after rescaling with a conformal
factor $\Theta$, $\widetilde g\mapsto \Theta^2  \tilde g=: g$, it admits a conformal boundary $\scri^{\pm}$,
with $\Theta|_{\scri}=0$ and $\mathrm{d}\Theta|_{\scri}\ne 0$,
 through which the metric $g$
admits a \emph{smooth} extension.
His proposal has been subject to an extensive discussion in  the literature, in particular whether the notion of a \emph{polyhomogeneous} 
$\scri$ might be better suitable to cover all physical situations of interest (cf.\ \cite{christo, F_i0_2, F_17} and the references given there).
While the construction of Cauchy data which produce a polyhomogeneous $\scri$ has been recently accomplished \cite{hiva},
a characterization of data which yield a smooth $\scri$ is still open.
Only in a setting where the data are stationary near spatial infinity the smoothness of $\scri$  has been established \cite{d_s, dain}.

To deal with this problem it is convenient to blow-up spatial infinity to a cylinder $I$ \cite{F_i0} which  runs into $\scri^{\pm}$ at the
critical sets $I^{\pm}\cong S^2$.
Generically, Cauchy data on some Cauchy surface $\Sigma$ which admit a smooth extension through $I^0:=I\cap \Sigma\cong S^2$
develop logarithmic terms at $I^{\pm}$ \cite{F_i0, kroon0}, and it is expected that these log-terms spread over to $\scri^{\pm}$ and produce
a polyhomogeneous  rather than a smooth $\scri^{\pm}$.

Since  the conformal field equations reduce to inner equations on the cylinder $I$ all fields and their derivatives can be computed on $I$ in terms of the Cauchy data.
However, as this requires integrations, it is  hard to control the fields at any order at $I^{\pm}$, which is necessary to analyze the appearance of log-terms.
For this reason an  asymptotic Cauchy problem was considered in \cite{ttp_i0}
where the data are prescribed on (a portion of) $\scri^-$ and some, to a large extent irrelevant, incoming null hypersurface.
As then the critical set $I^-$ is part of the initial surface (more precisely it arises as its limit) the appearance of log-terms and the corresponding restrictions on the data
can be much better understood.

In both scenarios it turns out to be very convenient to use a conformal Gauss gauge \cite{F_AdS, F_i0} which is based on a congruence of (timelike)
conformal geodesics. This comes along with the freedom to specify certain initial data for the conformal geodesics.
For the ordinary Cauchy problem a natural choice is to choose them orthogonal to $\Sigma$. For the asymptotic Cauchy problem such a natural
choice is not available. Instead, the choice of these gauge data was guided by the Minkowski case, together with the aim to simplify the analysis
at $I^-$ as much as possible.

The purpose of this note is to gain some insight how these two gauges are related at $I^-$.
Somewhat unexpectedly, it turns out that they belong to two quite different classes of gauges. Their relation is rather  complicated at $I^-$ and, in fact,
comes along with a gauge transformation
which is  polyhomogeneous there. This observation seems to make it very hard to translate smoothness results obtained in one gauge into the other.

This note is organized as follows: 
In Section~\ref{sec_cfe_cg} we review the conformal field equations and the conformal Gauss gauge.
In Section~\ref{sec_Schwarzschild} we will work out the gauge data which need to be prescribed on $\scri^-$ for both
types of conformal Gauss gauges by considering, as an example, the Schwarzschild metric.
 We generalize in Section~\ref{sec_comparison} the features worked out in  Section~\ref{sec_Schwarzschild}.
We then derive the no-logs conditions at $I^-$ and analyze the appearance of logarithmic terms in both gauges.
In Section~\ref{sec_trans_I} we also compute a conformal Gauss gauge for which the solutions are manifestly time-symmetric up to and including
the Schwarzschild order.

By restricting  attention to the flat case, we study the behavior of the gauge transformation
which connects both classes of conformal Gauss gauges in  Section~\ref{sec_gauge}.
Finally, in Section~\ref{sec_deviation_equation} we attempt to extract a geometric feature which distinguishes both gauges
by considering the deviation equation for a congruence of conformal geodesics.

\section{Conformal field equations and conformal Gauss gauge}
\label{sec_cfe_cg}

\subsection{General conformal field equations}

The  \emph{conformal field equations (CFE)} \cite{F1,F2}  substitute 
Einstein's vacuum field equations in Penrose's conformally rescaled  spacetimes. They are equivalent to the vacuum equations
in regions where the conformal factor does not vanish, and  remain regular  at points where it vanishes.
Beside the usual gauge freedom arising from the freedom to choose coordinates and  frame field, the CFE  contain an additional gauge freedom which arises from the artificially introduced conformal factor $\Theta$.

In fact, it turns out to be very convenient to introduce additional gauge degrees of freedom, which even more exploit the conformal structure. They are obtained when replacing the Levi-Civita connection
by some appropriately chosen Weyl connection. This way one is led to the so-called \emph{general conformal field equations (GCFE)},
introduced by Friedrich in \cite{F_AdS}, cf.\ \cite{F_i0, F3, F_i0_2}.

Let $(\widetilde\mcM, \widetilde g)$ be a smooth
Lorentzian manifold, and denote by $g=\Theta^2 \widetilde g$ its conformally rescaled counterpart.
We denote by $\nabla$ the Levi-Civita connection of  $g$.

Let $ f$ be a smooth 1-form on $\mcM$.
There exists a unique 
torsion-free connection $\widehat\nabla$,  the  so-called \emph{Weyl connection}, which satisfies
\begin{equation}
\widehat\nabla_{\sigma}  g_{\mu\nu} = -2  f_{\sigma}  g_{\mu\nu} \,.
\label{weyl_relation}
\end{equation}
Then
\begin{equation}
\widehat\nabla = \nabla + S( f)\,, \quad \text{where}\quad S(   f)_{\mu}{}^{\sigma}{}_{\nu}:=2\delta_{(\mu}{}^{\sigma}  f_{\nu)} -  g_{\mu\nu} g^{\sigma\rho} f_{\rho}
\,.
\end{equation}

Let $e_k$ be a frame field satisfying  $g(e_i,e_j)=\eta_{ij}\equiv \mathrm{diag}(-1,1,1,1)$.
We define the connection coefficients of $\widehat \nabla$ in this frame field by
\begin{equation}
\widehat\nabla_i e_j = \widehat\Gamma_i{}^k{}_j e_k
\,.
\end{equation}
They are related to the  connection coefficients of the Levi-Civita connection as follows,
\begin{equation}
 \widehat\Gamma_i{}^k{}_j =  \Gamma_i{}^k{}_j  +S(f)_i{}^k{}_j \,, \quad \text{and} \quad 
f_i =\frac{1}{4}\widehat\Gamma_i{}^k{}_k
\,.
\end{equation}
Finally, we set
\begin{equation}
b:=\Theta f + \mathrm{d}\Theta
\,,
\label{dfn_d}
\end{equation}
and denote by
\begin{align}
\widehat W^{\mu}{}_{\nu\sigma\rho} =&  \Theta^{-1}\widehat C^{\mu}{}_{\nu\sigma\rho}
\,,
\\
\widehat L_{\mu\nu} =& \frac{1}{2}\widehat R_{(\mu\nu)} - \frac{1}{4} \widehat R_{[\mu\nu]} - \frac{1}{12}\widehat R g_{\mu\nu}
\,,
\end{align}
\emph{ rescaled Weyl tensor} and \emph{Schouten tensor} of $\widehat\nabla$, respectively.
We note that
\begin{equation}
\widehat L_{\mu\nu} =L_{\mu\nu} - \nabla_{\mu}f_{\nu}  + \frac{1}{2}S(f)_{\mu}{}^{\sigma}{}_{\nu} f_{\sigma}
\,,
\label{schouten_weylconnect}
\end{equation}
and that  the rescaled Weyl tensor does not depend on the Weyl connection,
\begin{equation}
\widehat W^{\mu}{}_{\nu\sigma\rho}=W^{\mu}{}_{\nu\sigma\rho}
\,.
\end{equation}

Let now $(\widetilde\mcM, \widetilde g)$ be a
solution to  Einstein's vacuum field equations  $\widetilde R_{\mu\nu}=0$.
Then the tuple $ (e^{\mu}{}_k, \widehat\Gamma_i{}^k{}_j, \widehat L_{ij}, W^i{}_{jkl})$,
where $e^{\mu}{}_k:=\langle\mathrm{d} x^{\mu},e_k\rangle$,
satisfies the \emph{general conformal field equations (GCFE)} \cite{F_AdS}
\begin{align}
[e_p,e_q] =&  2\widehat\Gamma_{[p}{}^l{}_{q]} e_l
\,,
\label{GCFE_1}
\\
e_{[p}(\widehat\Gamma_{q]}{}^i{}_j) =&
 \widehat\Gamma_k{}^i{}_j\widehat\Gamma_{[p}{}^k{}_{q]}
- \widehat\Gamma_{[p}{}^i{}_{|k|}\widehat\Gamma_{q]}{}^k{}_j
+ \delta_{[p}{}^i\widehat L_{q]j} - \delta_j{}^i \widehat L_{[pq]} - \eta_{j[p}\widehat L_{q]}{}^i +\frac{\Theta}{2}   W^i{}_{jpq}\,,
\\
2\widehat\nabla_{[k}\widehat L_{l]j} =& b_i W^{i}{}_{jkl}\,,
\label{GCFE_3}
\\
 \widehat\nabla_i W^i{}_{jkl}  =& \frac{1}{4}\widehat\Gamma_i{}^p{}_pW^i{}_{jkl} 
\label{GCFE_4}
\,.
\end{align}
The fields $\Theta$ and $b$ reflect the \emph{conformal gauge freedom}.

\subsection{Conformal geodesics and   conformal Gauss gauge}

\subsubsection{Conformal geodesics}

A \emph{conformal geodesic for $(\mcM ,  g)$}
(cf.\ e.g.\ \cite {F_cg, F_Schmidt})
 is a curve $x(\tau)$ in $\mcM$
for which   a 1-form $ f=  f(\tau)$ exists along $x(\tau)$ such that the pair $(x,  f)$ solves
the \emph{conformal geodesics equations}
\begin{align}
\nabla_{\dot x} \dot x  =& -S( f)(\dot x,\cdot, \dot x)
\,,
\label{f_eqn1}
\\
\nabla_{\dot x}  f  =& \frac{1}{2}  S( f)(\dot x,f,\cdot)+  L(\dot x,\cdot)
\label{f_eqn2}
\,,
\end{align}
Given data $x_*\in  \mcM$, $\dot x_*\in T_{x_*}\mcM$ and $ f_*\in T^*_{x_*}\mcM$
there exists a unique solution $x(\tau)$, $ f(\tau)$  to  \eq{f_eqn1}-\eq{f_eqn2} near  $x_*$
satisfying, for given $\tau_*\in\mathbb{R}$,
\begin{equation}
x(\tau_*) = x_*\,, \quad \dot x(\tau_*) =  \dot x_* \,, \quad  f (\tau_*) =  f_*
\,.
\end{equation}
Some properties of conformal geodesics are analyzed in \cite{F_cg, F_Schmidt}.

Let us consider now a congruence $\{ x(\tau,\rho), f(\tau,\rho)\}$ of conformal geodesics, set $x':=\partial x/\partial \rho$, and denote by $F:= \nabla_{x'}f$ the \emph{deviation 1-form}.
The \emph{conformal Jacobi equation} reads \cite{F_cg}
\begin{equation}
a:=\nabla_{\dot x}\nabla_{\dot x} x'=\mathrm{Riem}(\dot x, x')\dot x - S(F)(\dot x, .,\dot x)
-2 S(f)(\dot x,.,\nabla_{\dot x}x')
\,.
\label{conf_dev}
\end{equation}
Such a tensor measures the relative acceleration of infinitesimally nearby geodesics in the family \cite{wald}.

\subsubsection{Conformal Gauss gauge}

It is convenient to introduce \emph{conformal Gauss coordinates} as a geometrically defined coordinate system, 
where the timelike coordinate lines are generated by timelike conformal geodesics.
Because of the additional terms in \eq{conf_dev} it is expected    that curvature induced tendencies to develop caustics (which often arise when dealing with  metric geodesics) may be counteracted \cite{F_cg}.

A \emph{conformal Gauss gauge} \cite{F_AdS,  F_i0} is  adapted to a congruence of (timelike) conformal geodesics, and  employs the fact that such a congruence  distinguishes the Weyl connection associated to the   1-form $f$.
It is constructed from an initial surface $\mathcal{S}$  (spacelike or null)  which intersects the congruence transversally and meets
each of the curves exactly once. 

One also has the gauge freedom to choose the conformal factor $\Theta$, on which we impose the condition that 
$\widehat \nabla$ preserves the conformal structure, i.e.\  
\begin{equation}
\widehat\nabla_{\dot x} g =0 \quad \Longleftrightarrow \quad 
 \langle \dot x , f \rangle =0
\,.
\label{conf_gauge1}
\end{equation}
The (eventually orthonormal) frame field $e_i$ will be parallely propagated w.r.t.\ $\widehat\nabla$,
\begin{equation}
\widehat\nabla_{\dot x} e_k=0
\,.
\label{conf_gauge2}
\end{equation}
This leaves the freedom to specify certain \emph{gauge data} on the initial surface.

In the case of an ordinary Cauchy problem with Cauchy surface $\Sigma$ \cite{F_i0} one can identify
\begin{equation}
\dot x|_{\Sigma}\,, \quad  f_{ \Sigma} \,, \quad \Theta|_{\Sigma}>0\;, \quad \widehat\nabla_{\dot x}\Theta|_{\Sigma}
\,, \quad \text{with}  \quad   g (\dot x, \dot x)|_{\Sigma}=-1
\,,
\label{F_gauge_data}
\end{equation}
together with a frame field $e_i$ on $\Sigma$ which satisfies
\begin{equation}
g(e_{i},e_{j})|_{\Sigma}=\eta_{ij} \quad (\Longrightarrow \quad g(e_{i},e_{j})=\eta_{ij})
\,.
\end{equation}
Here $  f_{\Sigma}$ denotes the pull back of  $ f$ on $\Sigma$.

In the case where the conformal geodesics are  constructed from $\scri^-$, which we assume to be of the ``natural'' topology
\begin{equation}
\scri^-\cong \mathbb{R}\times S^2
\,,
\end{equation}
one can identify the following gauge data \cite{ttp_i0},
\begin{align}
\dot x|_{\scri^-}\,, \quad  f_{ \scri^- } \,, \quad \widehat \nabla_{\dot x}\Theta|_{\scri^-}>0
 \;, \quad \theta^-
\quad \text{with}  \quad   g (\dot x, \dot x)|_{\Sigma} =-1
\,,
\label{rel_gauge_data}
\end{align}
where $\theta^-$ denotes the divergence  in the transverse direction, cf.\ \eq{dfn_theta-} below.
Equivalently, in adapted null coordinates \eq{adapted_null_gen} one may prescribe
\begin{align}
g_{\tau\tau}|_{\scri^-}=-1\,, \quad\nu_{\tau}>0\,, \quad \nu_{\mathring A}\,, \quad  f_{ \scri^- } \,, \quad \widehat \nabla_{\dot x}\Theta|_{\scri^-}>0
 \;, \quad \theta^-
\,,
\label{rel_gauge_data_alt}
\end{align}
where $g_{\tau\tau}|_{\scri^-}=-1$ arises from the requirement that  $g (\dot x, \dot x)|_{\Sigma}|_{\scri^-} =-1$.

Finally, \emph{conformal Gauss coordinates} are  obtained as follows:
One chooses  $x^0 =\tau $ together with local coordinates $x^{\alpha}$  on the initial surface which are 
then dragged along the conformal geodesics.
The parameter $\tau$  along the conformal geodesics will be  chosen in such a way that  $\Sigma=\{\tau=0\}$ and  $\scri^- =\{ \tau =-1\}$, respectively.

For given  ``conformal gauge data''
\eq{F_gauge_data} and \eq{rel_gauge_data} 
at least locally
a  gauge which satisfies \eq{f_eqn1}-\eq{f_eqn2}, \eq{conf_gauge1}, and \eq{conf_gauge2}
can be constructed \cite{ F_i0, ttp_i0}.
Coordinates, frame field, and conformal factor obtained  this way are said to form a \emph{conformal Gauss gauge}.
In such a gauge  the following relations are fulfilled,
\begin{equation}
\dot x= e_0=\partial_{\tau} \,, \quad g(e_i,e_j)=\eta_{ij}\,, \quad \hat L_{0k}=0\,, \quad \hat\Gamma_0{}^k{}_j =0
\,.
\label{0_gauge_conditions}
\end{equation}

\subsubsection{Coordinates and frame field}

On $\Sigma$ we choose coordinates $(x^0, x^{\alpha})$ such that
\begin{equation}
g|_{\Sigma}= -\mathrm{d}\tau^2 + h_{\alpha\beta}\mathrm{d}x^{\alpha}\mathrm{d}x^{\beta}
\,.
\end{equation}
As initial frame field we take
\begin{align}
e_0|_{\Sigma}=& \partial_{\tau}
\,,
\label{frame_Cauchy}
\\
e_a|_{\Sigma}=&{}^{(3)}e^{\alpha}{}_a\partial_{\alpha}
\,,
\label{frame_Cauchy2}
\end{align}
where ${}^{(3)}e_a$, $a=1,2,3$, denotes an orthonormal frame field on $(\Sigma, h)$.

On $\scri^-\cong \mathbb{R}\times S^2$  it  is  convenient to  take \emph{adapted null coordinates} $(\tau, r, x^{\mathring A})$,
which we will briefly introduce now.
They are defined in such a way that $\scri^-=\{\tau=-1\}$, $r$ parameterizes the null geodesic generators of $\scri^-$,  and the $x^{\mathring A}$'s are local coordinates
on the $\Sigma_r:=\{\tau=-1, r=\mathrm{const.}\}\cong S^2$-level sets (cf.\ \cite{CCM2} for more details).
In these coordinates the metric  adopts the form
\begin{equation}
 g|_{\scri^-}= - \mathrm{d}\tau ^2 + 2\nu_{\tau}(r,x^{\mathring C}) \mathrm{d}\tau\mathrm{d}r 
+ 2\nu_{\mathring A }(r,x^{\mathring C}) \mathrm{d}\tau\mathrm{d}x^{\mathring A}+\Omega^2(r,x^{\mathring C})  s_{\mathring A\mathring B}(x^{\mathring  C}) \mathrm{d}x^{\mathring A}\mathrm{d}x^{\mathring B}
\,,
\label{adapted_null_gen}
\end{equation}
where $s_{\mathring A\mathring B}$ denotes the standard metric on $S^2$ (note for this that the shear tensor vanishes on $\scri^-$ and that any
Riemannian metric on $S^2$ is conformal to the round metric).
Here and henceforth we use $\mathring{}$ to denote angular coordinate indices.
The  coefficients which appear in \eq{adapted_null_gen}  are determined by the constraint equations and also depend on how the coordinates are extended off $\scri^-$.

 At each $p\in \Sigma_r$ we denote by $\ell^{\pm}$ the future-directed null vectors orthogonal to $\Sigma_r$
and normalized in such a way that $ g(\ell^+,\ell^-) = -2$. In adapted null coordinates they read
\begin{equation}
 \ell^+=\partial_r\;, \quad \ell^- = -2\nu^{\tau}\partial_{\tau} -  g^{rr}\partial_r -2  g^{r\mathring A}\partial_{\mathring A}
\;,
\label{ell_eqn}
\end{equation}
where $\nu^{\tau}:=\nu_{\tau}^{-1}$.
We  denote by $\theta^{\pm}$ the \emph{divergences} of the null hypersurfaces emanating from $\Sigma_r$ tangentially to $\ell^{\pm}$.
We have
\begin{align}
 \theta^+(r,x^{\mathring A}) &\equiv [ g^{\mu\nu} + (\ell^+)^{(\mu} (\ell^-)^{\nu)}]\nabla_{\mu}\ell^+_{\nu}|_{\Sigma_r}
\nonumber
\\
 &= \frac{1}{2}g^{\mathring A\mathring B} \partial_r g_{\mathring A\mathring B}
\;,
\label{dfn_theta+}
\\
\theta^-(r,x^{\mathring A}) &\equiv  [ g^{\mu\nu} + (\ell^+)^{(\mu} (\ell^-)^{\nu)}]\nabla_{\mu}\ell^-_{\nu}|_{\Sigma_r}
\nonumber
\\
 &=2\nu^{\tau}  \rnabla^{\mathring A}\nu_{\mathring A} -\theta^+  g^{rr}  -\nu^{\tau}  g^{\mathring A\mathring B}\partial_{\tau}g_{\mathring A\mathring B}
\;,
\label{dfn_theta-}
\end{align}
where $\not\hspace{-.25em}\nabla$ denotes the Levi-Civita connection associated to the one-parameter family $r\mapsto \not\hspace{-.2em} g =g_{\mathring A\mathring B}\mathrm{d}x^{\mathring A}\mathrm{d}x^{\mathring B}|_{\scri^-}$ on $S^2$.

Adapted null coordinates come along with another gauge freedom, namely to reparameterize  the null geodesic generators of $\scri^-$.
This gauge freedom, $r\mapsto \tilde r=\tilde r(r, x^{\mathring A})$, can be employed to prescribe the function $\kappa$ \cite{CCM2},
given by
\begin{equation}
\rnabla_{\ell^+}\ell^+=\kappa\ell^+\,,
\end{equation}
which  measures the deviation of the coordinate $r$ to be an affine parameter.

As initial frame field we choose
\begin{align}
e_{0} |_{\scri^-}=& \partial_{\tau}
\,,
\label{frame_field1}
\\
e_{1} |_{\scri^-} =& \partial_{\tau} + \nu^{\tau}\partial_r
\,,
\label{frame_field2}
\\
e_{A}  |_{\scri^-}=&   \Omega^{-1}\mathring e^{\mathring A}{}_A( \partial_{\mathring A} - \nu^{\tau}\nu_{\mathring A}\partial_r )
\,,
\label{frame_field3}
\end{align}
where $(\mathring e_A)$, $A=2,3$, denotes an orthonormal frame field on the  round sphere $\mathbb{S}^2:=(S^2, s_{\mathring A\mathring B})$.

\subsubsection{Some crucial relations}

The conformal field equations  have been formulated in terms of the gauge fields $\Theta$ and $b\equiv \Theta f + \mathrm{d}\Theta$.
These are determined by the gauge conditions \eq{f_eqn1}, \eq{f_eqn2}, and  \eq{conf_gauge1} which, expressed in terms of
$\Theta$ and  $b$,  read
\begin{equation}
\widehat\nabla_{\dot x} \dot x  =0
\,,
\quad
 \widehat L(\dot x, \cdot) =0
\,,
\quad
\widehat\nabla_{\dot x}\Theta =\langle \dot x, b\rangle \,.
\label{eqn_Theta}
\end{equation}
%
A very remarkable result by Friedrich \cite{F_AdS} shows that the fields $\Theta$ and  $b$ can be explicitly determined in a conformal Gauss gauge,
so that the corresponding expressions can  be simply inserted into the conformal field equations.

\begin{lemma}[\cite{F_AdS}]
\label{lemma_b_theta}
In the conformal Gauss gauge the following relations hold:
\begin{enumerate}
\item[(i)] $\nabla_{\dot x}\nabla_{\dot x}\nabla_{\dot x}\Theta = 0$, and
\item[(ii)] $ \nabla_{\dot x}\nabla_{\dot x}b_k  = 0$, where $b_k\equiv \langle b, e_k\rangle$.
\end{enumerate}
\end{lemma}

Part of the data for (i) are provided by the gauge data on the initial surface. The remaining ones 
can be computed from the conformal field equations.

Viewed from $\Sigma$ we have
\begin{align}
\Theta =&\Theta^{(0)}+   \Theta^{(1)}\tau + \Theta^{(2)}\tau^2
\,,
\label{Theta_Sigma}
\\
b_i =& b^{(0)}_i + b^{(1)}_i \tau
\,,
\end{align}
with 
\cite{F_AdS} (recall \eq{frame_Cauchy}-\eq{frame_Cauchy2})
\begin{align}
\Theta^{(2)}  =&
-\frac{1}{4}\Theta^{(0)} f^a f_{a}
+\frac{1}{4}(\Theta^{(0)})^{-1} \Big((\Theta^{(1)})^2-|{}^{(3)}\nabla\Theta^{(0)}|^2\Big)
-\frac{1}{2}f^a{}^{(3)}\nabla_{a}\Theta^{(0)}
\,,
\\
b^{(0)}_0=&\Theta^{(1)}
\,,
\quad
b^{(0)}_a=\Theta^{(0)}f_a+ {}^{(3)}\nabla_a\Theta^{(0)}
\,,
\\
b_0^{(1)}
 =&-\frac{1}{2}( \Theta^{(0)})^{-1} \eta^{\sharp}(b^{(0)} ,b^{(0)})
\,,
\quad
b_a^{(1)}
 =0
\,.
\label{Theta_Sigma5}
\end{align}
Viewed from $\scri^-$ we have
\begin{align}
\Theta =&  \Theta^{(1)}(1+\tau) + \Theta^{(2)}(1+\tau)^2
\,,
\\
b_i =& b^{(0)}_i + b^{(1)}_i (1+\tau)
\,,
\end{align}
with \cite{ttp_i0}
\begin{align}
\Theta^{(2)}  =&
- \frac{1}{2} \Big(\rnabla_{\ell^+}+\kappa+ \langle \ell^+, f\rangle
\Big)\Big(\frac{\Theta^{(1)}}{  g(\dot x_*, \ell^+)}\Big)
\,,
\\
b^{(0)}_0=&\Theta^{(1)}
\,,
\quad
b^{(0)}_1= \Theta^{(1)}
\,,
\quad
b^{(0)}_A= 0
\,,
\\
b^{(1)}_0=& 2\Theta^{(2)}
\,,
\quad
b^{(1)}_1= 0
\,,
\quad
b^{(1)}_A= 0
\,.
\end{align}

\subsection{Transport equations}

The Levi-Civita connection satisfies $\Gamma_{i(jk)}=0$, equivalently, $\widehat\Gamma _{i(jk)} = \eta_{jk}f_i$. If follows that the Weyl connection
has the following (anti-)symmetric properties, we will make extensively use of,
\begin{eqnarray}
\widehat\Gamma_a{}^1{}_0=\widehat\Gamma_a{}^0{}_1\,, \quad \widehat\Gamma_a{}^0{}_0=\widehat\Gamma_a{}^1{}_1=\frac{1}{2}\widehat\Gamma_a{}^A{}_A\,, \quad\eta_{AB}\widehat\Gamma_a{}^B{}_1=-\widehat\Gamma_a{}^1{}_A\,, 
\quad \eta_{AB}\widehat\Gamma_a{}^B{}_0=\widehat\Gamma_a{}^0{}_A\,.
\label{relations_connection1}
\end{eqnarray}
As the relevant independent components
 of the Weyl connection one may regard
\begin{equation}
\widehat\Gamma_a{}^0{}_b
\,,\quad
\widehat\Gamma_a{}^1{}_b
\,,\quad
\widehat\Gamma_{a[BC]}
\,.
\end{equation}
As the independent components of the rescaled Weyl tensor it is convenient to identify
($(.)_{\mathrm{tf}}$ denotes the trace-free part w.r.t.\  the $(AB)$-components and $\eta_{AB}$)
\begin{equation}
W_{0101}\,, \quad W_{01AB}\,,\quad W^{\pm}_A:=W_{010A}\pm W_{011A} \,,\quad V^{\pm}_{AB}:= (W_{1A1B})_{\mathrm{tf}}\pm W_{0(AB)1}
\,.
\end{equation}

The GCFE imply the following evolution equations, whose evaluation on the cylinder $I$ (including radial derivatives)
provides the relevant transport equations (here we display a somewhat more explicit form as compared to \cite{ttp_i0})
for the Schouten tensor,
\begin{align}
\partial_{\tau}(\widehat L_{10} -\widehat L_{11} )
=&(b_0 -b_1)W_{0101}- \widehat\Gamma_1{}^1{}_0( \widehat L_{10}- \widehat L_{11} ) -  \widehat\Gamma_1{}^A{}_0(\widehat L_{A0} 
-\widehat L_{A1} )
\,,
\label{transport1}
\\
\partial_{\tau}(\widehat L_{10} +\widehat L_{11} )
=& -(b_0+b_1)W_{0101}-  \widehat\Gamma_1{}^1{}_0(\widehat L_{10}+\widehat L_{11} ) -  \widehat\Gamma_1{}^A{}_0(\widehat L_{A0} 
+\widehat L_{A1} )
\,,
\\
\partial_{\tau}(\widehat L_{A0} -\widehat L_{A1} )
=& \frac{1}{2}(b_0-b_1) (W^+_A+W^-_A)-  \widehat\Gamma_A{}^1{}_0(\widehat L_{10}-\widehat L_{11}) -  \widehat\Gamma_A{}^B{}_0(\widehat L_{B0} -\widehat L_{B1} )
\,,
\\
\partial_{\tau}(\widehat L_{A0} + \widehat L_{A1} )
=&- \frac{1}{2}(b_0+ b_1) (W^+_A+W^-_A)-  \widehat\Gamma_A{}^1{}_0(\widehat L_{10}+\widehat L_{11}) -  \widehat\Gamma_A{}^B{}_0(\widehat L_{B0} +\widehat L_{B1} )
\,,
\\
\partial_{\tau}\widehat L_{1A} 
=& -\frac{1}{2}(b_0-b_1) W^+_A- \frac{1}{2}(b_0+b_1)W^-_A
-  \widehat\Gamma_1{}^1{}_0\widehat L_{1A} 
-  \widehat\Gamma_1{}^B{}_0\widehat L_{BA} 
\,,
\\
\partial_{\tau}\widehat L_{AB} 
=& - \frac{1}{2}(b_0+b_1)V^+_{AB}- \frac{1}{2} (b_0-b_1)V^-_{AB}
+ \frac{1}{2}b_0 W_{0101}\eta_{AB} +\frac{1}{2}  b_1 W_{01AB}
\nonumber
\\
&
-  \widehat\Gamma_A{}^1{}_0\widehat L_{1B} -  \widehat\Gamma_A{}^C{}_0\widehat L_{CB} 
\,,
\end{align}
the connection coefficients,
\begin{align}
\partial_{\tau}\widehat\Gamma_{1}{}^0{}_1 
 =&
-  \widehat\Gamma_1{}^0{}_1\widehat\Gamma_{1}{}^1{}_{0} 
-  \widehat\Gamma_A{}^0{}_1\widehat\Gamma_{1}{}^A{}_{0} 
+ \widehat L_{11}   -\Theta W_{0101}
\,,
\\
\partial_{\tau}\widehat\Gamma_{1}{}^1{}_1 
 =&
-  \widehat\Gamma_1{}^1{}_1\widehat\Gamma_{1}{}^1{}_{0}
-  \widehat\Gamma_A{}^1{}_1\widehat\Gamma_{1}{}^A{}_{0}
 +\widehat L_{10}   
\,,
\\
\partial_{\tau}(\widehat\Gamma_{1}{}^0{}_A -\widehat\Gamma_{1}{}^1{}_A )
 =&
- ( \widehat\Gamma_1{}^0{}_A- \widehat\Gamma_1{}^1{}_A)\widehat\Gamma_{1}{}^1{}_{0} 
- ( \widehat\Gamma_B{}^0{}_A- \widehat\Gamma_B{}^1{}_A)\widehat\Gamma_{1}{}^B{}_{0} 
+ \widehat L_{1A} 
-\Theta W^+_A
\,,
\\
\partial_{\tau}(\widehat\Gamma_{1}{}^0{}_A + \widehat\Gamma_{1}{}^1{}_A )
 =&
-(  \widehat\Gamma_1{}^0{}_A+ \widehat\Gamma_1{}^1{}_A)\widehat\Gamma_{1}{}^1{}_{0} 
-  (\widehat\Gamma_B{}^0{}_A+ \widehat\Gamma_B{}^1{}_A)\widehat\Gamma_{1}{}^B{}_{0} 
+ \widehat L_{1A}  
-\Theta W^-_A
\,,
\\
\partial_{\tau}\widehat\Gamma_{A}{}^0{}_1
 =&
-  \widehat\Gamma_1{}^0{}_1\widehat\Gamma_{A}{}^1{}_{0}
- \widehat\Gamma_B{}^0{}_1\widehat\Gamma_{A}{}^B{}_{0}
+ \widehat L_{A1}   -\frac{1}{2} \Theta (W^+_A+W^-_A)
\,,
\\
\partial_{\tau}\widehat\Gamma_{A}{}^1{}_1 
 =&
-  \widehat\Gamma_1{}^1{}_1\widehat\Gamma_{A}{}^1{}_{0}
-\widehat\Gamma_B{}^1{}_1\widehat\Gamma_{A}{}^B{}_{0}
 +\widehat L_{A0}    
\\
\partial_{\tau}(\widehat\Gamma_{A}{}^0{}_B -\widehat\Gamma_{A}{}^1{}_B )
 =&
- ( \widehat\Gamma_1{}^0{}_B-\widehat\Gamma_1{}^1{}_B)\widehat\Gamma_{A}{}^1{}_{0} 
-  (\widehat\Gamma_C{}^0{}_B-  \widehat\Gamma_C{}^1{}_B)\widehat\Gamma_{A}{}^C{}_{0} 
+ \widehat L_{AB}   
\nonumber
\\
&
-\Theta V^-_{AB}
+\frac{1}{2}\Theta W_{0101}\eta_{AB}   -\frac{1}{2}\Theta W_{01AB}
\,,
\\
\partial_{\tau}(\widehat\Gamma_{A}{}^0{}_B + \widehat\Gamma_{A}{}^1{}_B )
 =&
- ( \widehat\Gamma_1{}^0{}_B+  \widehat\Gamma_1{}^1{}_B)\widehat\Gamma_{A}{}^1{}_{0} 
- ( \widehat\Gamma_C{}^0{}_B+ \widehat\Gamma_C{}^1{}_B)\widehat\Gamma_{A}{}^C{}_{0} 
+ \widehat L_{AB}  
\nonumber
\\
&
-\Theta V^+_{AB}
+\frac{1}{2}\Theta W_{0101}\eta_{AB}  -\frac{1}{2}\Theta W_{01AB}
\,,
\\
\partial_{\tau}\widehat\Gamma_{1}{}^A{}_B 
 =&
-  \widehat\Gamma_1{}^A{}_B\widehat\Gamma_{1}{}^1{}_{0} 
-  \widehat\Gamma_C{}^A{}_B\widehat\Gamma_{1}{}^C{}_{0} 
+ \delta^A{}_{B}\widehat L_{10}   +\Theta W^A{}_{B01}
\,,
\\
\partial_{\tau}\widehat\Gamma_{A}{}^B{}_C 
 =&
-  \widehat\Gamma_1{}^B{}_C\widehat\Gamma_{A}{}^1{}_{0} 
 -  \widehat\Gamma_D{}^B{}_C\widehat\Gamma_{A}{}^D{}_{0} 
+ \delta^B{}_{C}\widehat L_{A0}   -\Theta \eta_{A[B}(W^+_{C]}-W^-_{C]})
\,,
\end{align}
and the frame coefficients,
\begin{align}
\partial_{\tau}e^{\tau}{}_1=& -\widehat\Gamma_{1}{}^0{}_{0} 
 -\widehat\Gamma_{1}{}^1{}_{0} e^{\tau}{}_1
 -\widehat\Gamma_{1}{}^A{}_{0} e^{\tau}{}_A
\,,
\\
\partial_{\tau}e^{r}{}_1=&
 -\widehat\Gamma_{1}{}^1{}_{0} e^{r}{}_1
 -\widehat\Gamma_{1}{}^A{}_{0} e^{r}{}_A
\,,
\\
\partial_{\tau}e^{\mathring A}{}_1=& 
 -\widehat\Gamma_{1}{}^1{}_{0} e^{\mathring A}{}_1
 -\widehat\Gamma_{1}{}^A{}_{0} e^{\mathring A}{}_A
\,,
\\
\partial_{\tau}e^{\tau}{}_A=& -\widehat\Gamma_{A}{}^0{}_{0}
 -\widehat\Gamma_{A}{}^1{}_{0} e^{\tau}{}_1
 -\widehat\Gamma_{A}{}^B{}_{0} e^{\tau}{}_B
\,,
\\
\partial_{\tau}e^r{}_A=&
 -\widehat\Gamma_{A}{}^1{}_{0} e^{r}{}_1
 -\widehat\Gamma_{A}{}^B{}_{0} e^{r}{}_B
\,,
\\
\partial_{\tau}e^{\mathring A}{}_A=& 
 -\widehat\Gamma_{A}{}^1{}_{0} e^{\mathring A}{}_1
 -\widehat\Gamma_{A}{}^B{}_{0} e^{\mathring A}{}_B
\,.
\label{transport22}
\end{align}
The corresponding equations for the rescaled Weyl tensor are given in a very explicit form in \cite[eqns (2.90)-(2.99)]{ttp_i0}
and \cite[eqns (5.38)-(5.43)]{ttp_i0}.

\section{Two different representations of the Schwarzschild metric}
\label{sec_Schwarzschild}

In \cite{ttp_i0} we have  introduced a ``weakly asymptotically Minkowski-like conformal Gauss gauge''. By considering the Schwarzschild metric
as an example we want to work out the differences between this gauge and the one used by Friedrich \cite{F_i0}
to represent the Schwarzschild line element.

\subsection{Friedrich's gauge}

On an appropriate Cauchy surface $\Sigma$ in the conformally rescaled spacetime $(\mcM, g)$ the initial data
for the Schwarzschild metric are provided by the first and second fundamental form,
\begin{align*}
h=\Big(1+ \frac{m r}{2}\Big)^{-2}\Big(r^{-2}\mathrm{d}r^2+ s_{\mathring A\mathring B}\mathrm{d}x^{\mathring A}\mathrm{d}x^{\mathring B}\Big)
\,,\quad \chi=0
\,.
\end{align*}
As ``gauge data''  Friedrich took (similar to \eq{rel_gauge_data}-\eq{rel_gauge_data_alt}  this is equivalent to \eq{F_gauge_data})
\begin{align*}
\Theta|_{\Sigma}
=&r\Big(1+\frac{mr}{2}\Big)^{-3}
\,, \quad
\nabla_{\dot x}\Theta|_{\Sigma}= 0
\,,
\\
e_0|_{\Sigma}=&\partial_{\tau}
\,,
\quad
e_1|_{\Sigma} =r\Big(1+ \frac{m r}{2}\Big)  \partial_r
\,,
\quad
e_A |_{\Sigma}=\Big(1+ \frac{m r}{2}\Big) \mathring e_A
\,,
\\
f_0|_{\Sigma}=&0
\,,
\quad
f_1 |_{\Sigma}=1+ mr
\,,
\quad
f_A |_{\Sigma}=0
\,,
\\
g_{\tau\tau}|_{\Sigma}=&-1\,, \quad g_{\tau\alpha}|_{\Sigma}=0
\,,
\end{align*}
where $\mathring e$ denotes an orthonormal frame on the round sphere.

First of all we observe by \eq{Theta_Sigma}-\eq{Theta_Sigma5}
that this implies that, globally,
\begin{align*}
\Theta=&r \Big(1 + \frac{mr}{2}\Big)^{-3}\big(1-\tau^2\big)
\,,
\\
b_0=&- 2\tau r \Big(1 + \frac{mr}{2}\Big)^{-3}
\,,
\\
b_1 =&2 r\Big(1 + \frac{mr}{2}\Big)^{-3}
\,,
\\
b_A =&0
\,.
\end{align*}
%
%
From this we compute the connection coefficients on $\Sigma$. The non-vanishing ones are (up to those which are not independent of these ones)
\begin{align*}
\widehat\Gamma_1{}^1{}_1 |_{\Sigma}=&1+ mr
\,,
\\
\widehat\Gamma_A{}^1{}_B |_{\Sigma}=&  - \Big(1+\frac{mr}{2}\Big)\eta_{AB}
\,,
\\
\widehat\Gamma_A{}^B{}_C |_{\Sigma}=&\Big(1+ \frac{m r}{2}\Big) \mathring \Gamma_A{}^B{}_C 
\,.
\end{align*}
For the non-vanishing  components of the Schouten tensor we find
\begin{align*}
\widehat L_{11}|_{\Sigma} =& L_{11}+\frac{1}{2} =2 mr 
\,,
\\
\widehat L_{AB}|_{\Sigma} =& L_{AB} -\frac{1}{2}(1+mr)\eta_{AB}=-mr\eta_{AB}
\,,
\end{align*}
where $L_{ij}$ has been computed from the conformal field equation \cite{F3}
\begin{equation*}
\nabla_i\nabla_j\Theta=-\Theta L_{ij}  + \frac{1}{2\Theta}\nabla_k\Theta\nabla^k\Theta \eta_{ij}
\,.
\end{equation*}

Finally, for the only non-vanishing independent component of the rescaled Weyl tensor on $\Sigma$  we find
from the conformal field equations
\begin{equation}
W_{0101} |_{\Sigma} =-2m\Big(1+\frac{mr}{2}\Big)^3
\,.
\label{data_Weyl_I}
\end{equation}
while all transverse derivatives vanish on $\Sigma$,
\begin{equation}
\partial_{\tau}W_{ijkl} |_{\Sigma} =0
\,.
\label{trans_data_Weyl}
\end{equation}

Once we know the values of all the unknowns which appear in the GCFE on the Cauchy surface $\Sigma$ (including some of their transverse derivatives) we have all the initial data
at hand to solve the transport equation on the cylinder.
We will see that the  relevant features which distinguish this gauge from the one used in \cite{ttp_i0}  arise in the first-order radial derivatives on $I$.

Evaluation of the transport equations \eq{transport1}-\eq{transport22} on $I$ 
 yields (at this order one obtains the same as in \cite{ttp_i0}
and the same as for the Minkowski spacetime in its standard conformal representation)
\begin{align}
\widehat L_{ij} |_{I} 
=&0
\,,
\label{Mink_gauge1}
\\
\widehat\Gamma_{a}{}^0{}_b |_{I} 
 =&
0
\,,
\quad
\widehat\Gamma_{1}{}^i{}_j |_{I} 
 =
\delta^i{}_j
\,,
\\
\widehat\Gamma_{A}{}^1{}_1 |_{I} 
 =&
0
\,,
\quad
\widehat\Gamma_{A}{}^1{}_B |_{I} 
 =
-\eta_{AB}
\,,
\quad
\widehat\Gamma_{A}{}^B{}_C |_{I} 
 =
\mathring \Gamma_A{}^B{}_C 
\,,
\\
e^{\tau}{}_1 |_{I}  =& - \tau
\,,
\quad
e^{r}{}_1 |_{I}  =0
\,,
\quad
e^{\mathring A}{}_1 |_{I}  = 0
\,,
\\
e^{\tau}{}_A |_{I}  =& 0
\,,
\quad
e^{r}{}_A |_{I}  = 0
\,,
\quad
e^{\mathring A}{}_A |_{I}  = \mathring e^{\mathring A}{}_A
\,,
\end{align}
for Schouten tensor, frame and connection coefficients, while we find for the rescaled Weyl tensor the  values,
\begin{equation}
W_{0101}|_I=-2m\,, \quad W_{01AB}|_I=0\,,\quad W^{\pm}_A|_I=0 \,,\quad V^{\pm}_{AB}|_I=0
\,.
\label{data_Weyl_I}
\end{equation}
For the first-order radial derivatives we find
\begin{align*}
\partial_r \widehat L_{10} |_I
=&4m\tau
\,,
\quad
\partial_r\widehat L_{11}|_I 
=2m(1- \tau^2 )
\,,
\quad
\partial_r\widehat L_{1A}|_I 
=0
\,,
\\
\partial_r \widehat L_{A0} |_I
=&0
\,,
\quad
\partial_r\widehat L_{A1}|_I 
=0
\,,
\quad
\partial_r\widehat L_{AB}|_I 
=-m(1-\tau^2 )\eta_{AB} 
\,,
\\
\partial_r\widehat\Gamma_{1}{}^0{}_1 |_I
 =&
4m\Big(\tau- \frac{1}{3}\tau^3 \Big) 
\,,
\quad
\partial_r\widehat\Gamma_{1}{}^0{}_A |_I
 =0
\,,
\\
\partial_r\widehat\Gamma_{A}{}^0{}_1 |_I
 =&
0
\,,
\quad
\partial_r\widehat\Gamma_{A}{}^0{}_B |_I
 =
-2m\Big(\tau-\frac{1}{3}\tau^3 \Big)\eta_{AB}  
\,,
\\
\partial_r\widehat\Gamma_{1}{}^1{}_1 |_I
 =&
  m\Big(1+ \frac{1}{3} \tau^4\Big)
\,,
\quad
\partial_r\widehat\Gamma_{1}{}^1{}_A |_I
 =
0
\,,
\\
\partial_r\widehat\Gamma_{A}{}^1{}_1 |_I
 =&
0
\,,
\quad
\partial_r\widehat\Gamma_{A}{}^1{}_B |_I
 =
-m\Big(\frac{1}{2}+ \tau^2 -\frac{1}{6}\tau^4 \Big)\eta_{AB}  
\,,
\\
\partial_r(\widehat\Gamma_{1}{}^A{}_B)_{\mathrm{tf}} |_I
 =&
0
\,,
\quad
\partial_r\widehat\Gamma_{A}{}^B{}_C |_I
 =
m\Big(\frac{1}{2} + \tau^2-\frac{1}{6}\tau^4 \Big)\  \mathring \Gamma_A{}^B{}_C
\,,
\\
\partial_re^{\tau}{}_1 |_I =&  -  m\Big(\tau -\frac{4}{3}\tau^3+ \frac{1}{3} \tau^5\Big)
\,,
\quad
\partial_re^{r}{}_1 |_I = 
1
\,,
\quad
\partial_re^{\mathring A}{}_1 |_I =
0
\,,
\\
\partial_re^{\tau}{}_A |_I =& 0
\,,
\quad
\partial_re^{r}{}_A |_I = 
0
\,,
\quad
\partial_re^{\mathring A}{}_A |_I = 
 m\Big(\frac{1}{2}+ \tau^2-\frac{1}{6}\tau^4 \Big) e^{\mathring A}{}_A
\,.
\end{align*}
For the radial derivatives of the rescaled Weyl tensor we obtain,
\begin{align*}
(1-\tau) \partial_{\tau}\partial_r V^+_{AB}|_I
 =&
\partial_rV^+_{AB}
+ \Big(  \mcD_{(A} \partial_rW^-_{B)}
\Big)_{\mathrm{tf}}
\,,
\\
(1+\tau) \partial_{\tau}
 \partial_rV^-_{AB} |_I
=&
- \partial_rV^-_{AB} 
-\Big( \mcD_{(A}\partial_rW^+_{B)}
\Big)_{\mathrm{tf}}
\,,
\\
(1+\tau)\partial_{\tau}\partial_rW^-_{A}   |_I
 =&  2\mcD^B\partial_rV^+_{AB}
+2\partial_rW^-_{A}
\,,
\\
(1-\tau)\partial_{\tau}\partial_rW^+_{A} |_I
 =& -  2 \mcD^B \partial_rV^-_{AB}
- 2\partial_rW^+_{A}
\,,
\\
\partial_{r} W_{0101}|_I
 =& 
-\frac{1}{2}(1+\tau)\mcD^A \partial_rW^{+}_A
-\frac{1}{2}(1-\tau)\mcD^A\partial_rW^{-}_A
-m^2\Big(  3+ 6\tau^2-\tau^4   \Big)
\,,
\\
\partial_{r}W_{01AB} |_I   =&
(1+\tau)\mcD_{[A}   \partial_rW^{+}_{B]}
-(1-\tau)\mcD_{[A}  \partial_rW^{-}_{B]}
\,.
\end{align*}
This implies decoupled equations for $W^{\pm}_A$,
\begin{align*}
(1-\tau^2) \partial_{\tau}^2\partial_rW^-_{A} |_I 
 =&
  2\partial_{\tau}\partial_rW^-_{A}-2\partial_rW^-_{A}
+(\mathring\Delta+1) \partial_rW^-_{A}
\,,
\\
 (1-\tau^2)\partial_{\tau}^2\partial_rW^+_{A} |_I 
=&
-2\partial_{\tau}\partial_rW^+_{A} - 2\partial_rW^+_{A}
+(\mathring \Delta+1) \partial_rW^+_{A}
\,.
\end{align*}
The data \eq{data_Weyl_I} and \eq{trans_data_Weyl}
imply the trivial solutions, whence
\begin{align*}
\partial_rW^{\pm}_{A}  |_I=0
\,,
\quad
\partial_r V^{\pm}_{AB}|_I
 =
0
\,,
\quad
\partial_{r}W_{01AB} |_I   =0
\,,
\quad
\partial_{r} W_{0101}|_I
 =
-m^2\Big(  3+ 6\tau^2-\tau^4   \Big)
\,.
\end{align*}
We also compute those second-order radial derivatives on $I$ which are needed to determine
the  second-order radial derivatives of the frame coefficients on $I$ (except for $\partial_r^2e^{\mathring A}{}_A|_I$ which will not be needed),
\begin{align*}
\partial_r^2 \widehat L_{10}|_I
=&-\frac{4}{3}m^2(2\tau^3 -\tau^5)
\,,
\\
\partial_r^2 \widehat L_{11} |_I
=& -\frac{4}{3}m^2\Big(6\tau^2+\frac{1}{2}\tau^4+\frac{1}{6}\tau^6\Big)
\,,
\\
\partial_r^2\widehat L_{A0} |_I
=& 0
\,,
\quad
\partial_r^2 \widehat L_{A1} |_I
=0
\,,
\quad
\partial_r^2\widehat L_{1A} |_I
=
0
\,,
\\
\partial_r^2\widehat\Gamma_{1}{}^0{}_1 |_I
 =&
-\frac{4}{3}m^2\Big(7\tau^3 -\tau^5 + \frac{4}{21}\tau^7\Big)
\,,
\\
\partial_r^2\widehat\Gamma_{1}{}^1{}_1 |_I
 =&
-m^2\Big(4\tau^2-\frac{7}{3}\tau^4+\frac{4}{9}\tau^6-\frac{1}{7}\tau^8
\Big)
\,,
\\
\partial_r^2\widehat\Gamma_{1}{}^0{}_A |_I
 =&0
\,,
\quad
\partial_r^2\widehat\Gamma_{1}{}^1{}_A |_I
 =
0
\,,
\quad
\partial_r^2\widehat\Gamma_{A}{}^0{}_1 |_I
 =
0
\,,
\quad
\partial_r^2\widehat\Gamma_{A}{}^1{}_1  |_I
 =
0
\,,
\\
\partial_r^2e^{\tau}{}_1|_I=&m^2\Big(4\tau^3 -5 \tau ^5+\frac{8}{7}\tau^7-\frac{1}{7}\tau^9\Big)
\,,
\\
\partial_r^2e^{r}{}_1|_I =&
 m\Big(1-4\tau^2+ \frac{2}{3}\tau^4 \Big) 
\,,
\\
\partial_r^2e^{\mathring A}{}_1|_I=& 
0
\,,
\quad
\partial_r^2e^{\tau}{}_A|_I =0
\,,
\quad
\partial_r^2e^r{}_A|_I=
0
\,.
\end{align*}
From the frame coefficients we determine the Schwarzschild metric near $I$ in conformal Gauss coordinates based
on Friedrich's choice of the congruence of conformal geodesics,
\begin{equation*}
g^{\mu\nu} = e^{\mu}{}_i e^{\nu}{}_j \eta^{ij}
= -e^{\mu}{}_0 e^{\nu}{}_0
+  e^{\mu}{}_1 e^{\nu}{}_1 
+  \eta^{AB} e^{\mu}{}_A e^{\nu}{}_B
\,,
\end{equation*}
which yields
\begin{align*}
g^{\tau\tau}|_{I}=& (1-\tau^2)\Big( -1+ \frac{2}{3}   m\tau^2(3-\tau^2)r
+\frac{1}{63} m^2 \tau^2\big(63 - 357 \tau^2 +112 \tau^4-16 t^6\big)r^2\Big)
+ O(r^3)
\,,
\\
g^{\tau r}|_{I}=&   -\tau r -  m\Big(\frac{3}{2}\tau -\frac{10}{3}\tau^3+ \frac{2}{3} \tau^5\Big) r^2+ O(r^3)
\,,
\\
g^{\tau\mathring A}|_{I}=&O(r^3)
\,,
\\
g^{rr}|_{I}=&r^2 +   m\Big(1-4\tau^2+ \frac{2}{3}\tau^4 \Big) r^3+ O(r^4)
\,,
\\
g^{r\mathring A}|_{I}=& O(r^3)
\,,
\\
g^{\mathring A\mathring B}|_{I}=& \Big(1
+ m\big(1+ 2\tau^2-\frac{1}{3}\tau^4 \big)r\Big)  s^{\mathring A\mathring B}
+ O(r^2)
\,.
\end{align*}
From these expression we determine an expansions of the metric $g^{\sharp}|_{\scri^-}$ at $I^-$. Taking the inverse then gives,
\begin{align*}
g_{\tau\tau}|_{\scri^-}=&   -1 
\,,
\\
\nu_{\tau}|_{\scri^-}=& \frac{1}{r} + \frac{7}{6} m + O(r)
\,,
\\
\nu_{\mathring A}|_{\scri^-}=&O(r^2)
\,,
\\
g_{rr}|_{\scri^-}=&  O(r)
\,,
\\
g_{r\mathring A}|_{\scri^-}=& O(r^2)
\,,
\\
g_{\mathring A\mathring B}|_{\scri^-}=& 
\Big( 1-  \frac{8}{3} mr\Big) s_{\mathring A\mathring B}
+ O(r^2)
\,.
\end{align*}
We will also employ  that 
\begin{align*}
\partial_{\tau}g_{rr}|_{\scri^-} =& \frac{2}{r^2} + \frac{2m}{r}+ O(1)
\,,
\\
\partial_{\tau}g^{\mathring A\mathring B}|_{\scri^-}=& -\frac{8}{3}mr  s^{\mathring A\mathring B}
+ O(r^2)
\,.
\end{align*}
From these expansions we determine the expansions at $I^-$ of the gauge functions
\eq{rel_gauge_data_alt}
on $\scri^-$ which produce Friedrich's gauge,
\begin{align}
\nu_{\tau} =& \frac{1}{r} + \frac{7}{6} m + O(r)
\,,
\label{Fgauge1}
\\
\nu_{\mathring A} =&  O(r^2)
\,,
\\
\Theta^{(1)} = &2 r -3m r^2+ O(r^3)
\,,
\\
\kappa= &-\frac{2}{r}+\frac{4}{3} m + O(r)
\,,
\\
\theta^- =&-\frac{8}{3} m r^2+ O(r^3)
\,,
\\
f_1|_{\scri^-} =& 1+ \frac{4}{3}mr +O(r^2)
\,,
\\
f_A|_{\scri^-} =&O(r^2)
\,,
\\
g_{\mathring A\mathring B} |_{I^-} =& s_{\mathring A\mathring B}
\,.
\label{Fgauge8}
\end{align}

\subsection{Alternative gauge}

In \emph{Kerr-Schild Cartesian coordinates} the Schwarzschild line elements reads,
\begin{equation*}
\widetilde g
= -(\mathrm{d}y^0)^2+(\mathrm{d}y^1)^2 + (\mathrm{d}y^2)^2 +(\mathrm{d}y^3)^2 - \frac{2m}{R}
\ell\otimes\ell
\,,
\end{equation*}
where the 1-form 
\begin{equation*}
\ell := \mathrm{d}y^0 -\mathrm{d}R
\end{equation*}
satisfies  $\eta^{\sharp}(\ell,\ell) =0$.
The function $R$ is given by
%
\begin{equation*}
R:=\sqrt{(y^1)^2+(y^2)^2+(y^3)^2 }
\,.
\end{equation*}
We apply the  coordinate transformation $(y^{\mu})\mapsto (\tau, r, x^{\mathring A})$,
\begin{align*}
y^0= \frac{-\tau}{r(1-\tau^2)} 
\,,
\qquad 
y^1= \frac{-\sin\theta\cos\phi}{r(1-\tau^2)}
\,,
\qquad
y^2= \frac{-\sin\theta\sin\phi}{r(1-\tau^2)}
\,, \qquad 
y^3= \frac{-\cos\theta}{r(1-\tau^2)}
\,.
\end{align*}
%
%
and choose the  conformal factor,
\begin{equation*}
\Theta := r (1-\tau^2)
\,.
\end{equation*}
%
%
%
Altogether we then end up with the following conformal representation of the  Schwarzschild metric,
\begin{align}
 g 
=&  -\mathrm{d}\tau^2 -2\frac{\tau}{r}\mathrm{d}\tau\mathrm{d} r + \frac{1-\tau^2}{r^2}\mathrm{d}r^2 +  \mathrm{d}\Omega_2
\nonumber
\\
&
-2m r\frac{(1+\tau)^3}{1-\tau}\mathrm{d}\tau^2
+4m (1+\tau)^3 \mathrm{d}\tau\mathrm{d}r
-\frac{2m}{r} (1-\tau)(1+\tau)^3\mathrm{d}r^2
\,.
\label{Schwarzschild_metric_alt}
\end{align}
We want this to be conformal Gaussian coordinates in the leading order at $\scri^-$. From the
conformal geodesics equations \eq{f_eqn1}-\eq{f_eqn2}
we deduce that
\begin{equation*}
\Gamma^{\mu}_{\tau\tau}= -S(f)_{\tau}{}^{\mu}{}_{\tau}=-f^{\mu}
\,.
\end{equation*}
On $\scri^-$ this equation is satisfied if we choose $f_r|_{\scri^-}=1/r$ and $f_{\mathring A}|_{\scri^-}=0$.

This way we are led to choose the following gauge functions on $\scri^-=\{\tau=-1,r>0\}$,
\begin{align}
\nu_{\tau} =& \frac{1}{r} 
\,,
\label{Pgauge1}
\quad
\nu_{\mathring A} = 0
\,,
\\
\Theta^{(1)} = &2 r 
\,,
\\
\kappa= &-\frac{2}{r}
\,,
\quad
\theta^- =0
\,,
\\
f_1|_{\scri^-} =& 1
\,,
\quad
f_A|_{\scri^-} =0
\,,
\\
g_{\mathring A\mathring B} |_{I^-} =& s_{\mathring A\mathring B}
\,.
\label{Pgauge8}
\end{align}

\section{Comparison of both conformal gauges}
\label{sec_comparison}

The gauge \eq{Pgauge1}-\eq{Pgauge8} provides the ``simplest'' choice of the gauge functions on $\scri^-$. 
In the Minkowski case this choice leads to its standard  cylinder representation.
The gauge  \eq{Fgauge1}-\eq{Fgauge8} is more adapted to Schwarzschild an yields a 
conformal representation where the Schwarzschild metric is manifestly time-symmetric.
We want to figure out in which way these gauges are different and if so, what the characterizing feature is.

In the leading order at $I^-$ the gauge data coincide (and coincide with the ``Minkowskian'' values).
We further observe that the gauge \eq{Pgauge1}-\eq{Pgauge8} belongs to the more general class of
\emph{weakly asymptotically Minkowski-like conformal Gauss gauges} introduced in \cite{ttp_i0}.
It has been shown there that given a solution to the GCFE which  belongs to this class and is smooth at $I^-$,
the solution remains smooth when passing to any other  weakly asymptotically conformal Gauss gauge.
To prove that certain assumptions on the next-to-leading order terms were imposed, namely,
$\kappa^{(0)}=0$ and $\Theta^{(1,2)}-2\nu_{\tau}^{(0)}=0$.
An analysis of the argument given there (which we will reconsider more detailed below) shows that these assumptions can be somewhat weakened.
Also on  $\theta^-$ one can impose a weaker decay condition.
Accordingly, in this paper a \emph{weakly asymptotically Minkowski-like conformal Gauss gauge} will be a conformal Gauss gauge where
the gauge data on $\scri^-$ satisfy
\begin{align}
\nu_{\tau} =& \frac{1}{r}  + \nu_{\tau}^{(0)} + \mathfrak{O}(r)
\,,
\label{genPgauge1}
\\
\nu_{\mathring A} =&\nu_{\mathring A}^{(1)}r+  \mathfrak{O}(r^2)
\,,
\\
\Theta^{(1)} = &2 r + \Theta^{(1,2)} r^2 + \mathfrak{O}(r^3)
\,,
\\
\kappa= &-\frac{2}{r} + \kappa^{(0)} + \mathfrak{O}(r)
\,,
\\
\theta^- =& \theta^{-(2)} r^2+ \mathfrak{O}(r^3)
\,,
\\
f_1|_{\scri^-} =& 1 +f_1{}^{(1)}r +  \mathfrak{O}(r)
\,,
\\
f_A|_{\scri^-} =&f_A^{(1)}r + \mathfrak{O}(r^2)
\,,
\\
g_{\mathring A\mathring B} |_{I^-} =& s_{\mathring A\mathring B}
\,,
\label{genPgauge8}
\end{align}
together with the additional  \emph{``Minkowski-like'' gauge condition}
\begin{equation}
\Sigma:= \kappa^{(0)} + \Theta^{(1,2)}-2\nu^{(0)}_{\tau} =0 \quad\overset{\eq{crucial_constr}}{\Longleftrightarrow} \quad
\kappa^{(0)}=\theta^{+(0)}
\,.
\label{the_gauge_cond}
\end{equation}
 We say that  $f= \mathfrak{O}(r^n)$, $n\geq 0$, if it is a smooth function of $r$
and $x^{\mathring A}$, and if it Taylor expansion at $r=0$ starts with a term of $n$th-order.
We say that $f= \mathfrak{O}(r^{-n})$ if $r^n f= \mathfrak{O}(1)$.

It will become clear in the following that the decisive feature which distinguishes this gauge from Friedrich's gauge is that in the
latter case \eq{the_gauge_cond} is violated.

To gain  a better understanding of the impact of the condition \eq{the_gauge_cond}
 it is useful to determine all the fields which appear in the GCFE and their first-order radial derivatives on the cylinder.
To ease the computations let us impose certain  assumptions:
We are particularly interested in spacetimes which, in the leading order, behave similar to the Schwarzschild spacetime.
Let us therefore restrict attention to spacetimes where   the limit of the mass aspect to $I^-$ is constant
and where  the corresponding limit of the dual mass aspect is constant as well an therefore  vanishes, cf.\ \cite{ttp_i0},
%
\begin{equation}
M=\mathrm{const.}=: -m\,, \quad N=0
\,.
\label{const_mass_aspect}
\end{equation}
(The minus sign appears as we stick to the sign convention used for $M$ in \cite{ttp_i0}.)

When computing the behavior of the fields near $I^-$ one observes that the gauge data $ \nu_{\tau}^{(0)} $, $\nu_{\mathring A}^{(1)}$,
$ \Theta^{(1,2)}$, $ \kappa^{(0)} $, $ \theta^{-(2)} $, $f_1{}^{(1)}$ and $f_A^{(1)}$ enter at the same order as the mass $m$, while
the radiation field, angular momentum etc.\ only appear in higher orders. 
It therefore seems reasonable to choose the scalars on $I^-$ to be constant and the vectors to vanish (as in  Friedrich's Schwarzschild gauge, though
the  constants can take  arbitrary values  at this stage).
In that case we will speak of a
\emph{weakly asymptotically Schwarzschild-like conformal Gauss gauge}, which to summarize, is a conformal Gauss gauge with the following gauge data given
at $\scri^-$,
\begin{align}
\nu_{\tau} =& \frac{1}{r}  + \nu_{\tau}^{(0)} + \mathfrak{O}(r)
\,,
\label{genFgauge1}
\\
\nu_{\mathring A} =& \mathfrak{O}(r^2)
\,,
\\
\Theta^{(1)} = &2 r + \Theta^{(1,2)} r^2 + \mathfrak{O}(r^3)
\,,
\\
\kappa= &-\frac{2}{r} + \kappa^{(0)} + \mathfrak{O}(r)
\,,
\\
\theta^- =& \theta^{-(2)} r^2+ \mathfrak{O}(r^3)
\,,
\\
f_1|_{\scri^-} =& 1 +f_1{}^{(1)}r +  \mathfrak{O}(r)
\,,
\\
f_A|_{\scri^-} =& \mathfrak{O}(r^2)
\,,
\\
g_{\mathring A\mathring B} |_{I^-} =& s_{\mathring A\mathring B}
\,,
\label{genFgauge8}
\end{align}
where 
\begin{equation}
 \nu_{\tau}^{(0)}\,, \enspace \Theta^{(1,2)}\,, \enspace \kappa^{(0)} \,, \enspace \theta^{-(2)}\,, \enspace f_1{}^{(1)} \text{ are constant on $I^-$.}
\end{equation}
We emphasize that \eq{the_gauge_cond} does \emph{not} need to be satisfied.
If, though, \eq{the_gauge_cond} is fulfilled, then this also belongs to the class of  weakly asymptotically \emph{Minkowski-like} conformal Gauss gauges.

\subsection{Solution of the constraints on $\scri^-$}
\label{sec_field_on_scri}

When solving the constraint equations on $\scri^-$ it is convenient to introduce the field
\begin{equation}
\Xi_{\mathring A\mathring B}:=-2( \Gamma{}^r_{\mathring A\mathring B})_{\mathrm{tf}} =\nu^{\tau} (\partial_{\tau} g_{\mathring A\mathring B})_{\mathrm{tf}}
-2 \nu^{\tau}(\rnabla_{(\mathring A}\nu_{\mathring B)})_{\mathrm{tf}}
\end{equation}
which is, up to certain  integration functions on $I^-$, in one-to-one correspondence to the radiation field \cite{ttp_i0}.
It has been shown in  \cite{ttp_i0} that for the rescaled Weyl tensor to be bounded at $I^-$ we need to choose 
$\Xi_{AB}=O(r^2)$. In fact, in view of  \eq{const_mass_aspect} we are led to consider data of the form
\begin{equation}
\Xi_{ A B}=\mathfrak{O}(r^3)
\,.
\label{cond_Xi}
\end{equation}
This is because the function $N$ turns out to be one of the Hodge-decomposition scalars of $\Xi_{AB}$, while the other
one can be identified with a gauge freedom which arises from a gauge  freedom to reparameterize  the null geodesics
generating $\scri^-$, cf.\ Remark~\ref{rem_second_datum}, which  is not exploited by $\kappa$.

We note that the constraint equations on $\scri^-$ \cite[Equations (A.9)-(A.19)]{ttp_i0}
imply the following expansions
\begin{align}
\xi_A=& \mcD_{ A}\Big(\frac{1}{2}\theta^{+(1)}-\kappa^{(1)}\Big)r^2+\mathfrak{O}(r^3)
\,,
\\
\Omega=&1 + \frac{1}{2}\theta^{+(0)} r + \frac{1}{4}\Big(\theta^{+(1)}+\frac{1}{2}(\theta^{+(0)})^2\Big)r^2+\mathfrak{O}(r^3)
\,,
\\
\not \hspace{-0.2em} R=&\frac{2-2\Delta_s\log\Omega}{\Omega^2}
=2 -2\theta^{+(0)} r + \Big((\theta^{+(0)})^2 -\theta^{+(1)}\Big)r^2
-\frac{\Delta_s \theta^{+(1)}}{2}r^2
+ \mathfrak{O}(r^3)
\,,
\\
g^{rr}=&r^2- 2\nu_{\tau}^{(0)}r^3+  \mathfrak{O}(r^4)
\,,
\end{align}
with
\begin{align}
\theta^{+(0)}=& 2\kappa^{(0)}  + \Theta^{(1,2)}-2\nu^{(0)}_{\tau} 
\,.
\label{crucial_constr}
\end{align}
From \eq{frame_field2}-\eq{frame_field3} we find the following expansions for frame field,
\begin{align}
e_1|_{\scri^-}=& \partial_{\tau} +(r -\nu^{(0)}_{\tau}r^2+ \mathfrak{O}(r^3))\partial_r
\,,
\\
e_A|_{\scri^-} = & \Big(1-\frac{1}{2}\theta^{+(0)} r+  \mathfrak{O}(r^2)\Big)\mathring e^{\mathring A}{}_A\partial_{\mathring A}+ \mathfrak{O}(r^3)\partial_r
\,,
\end{align}
connection coefficients,
\begin{align*}
 \widehat\Gamma_A{}^1{}_1   |_{\scri^-}=&  \mathfrak{O}(r^2)
\,,
\\
 \widehat\Gamma_A{}^1{}_0   |_{\scri^-}=&  \mathfrak{O}(r^2)
\,,
\\
 \widehat\Gamma_A{}^0{}_B|_{\scri^-}  =& 
 - \frac{1}{4}\Big( \theta^{-(2)} + \theta^{+(0)}\Big)\eta_{AB}r+  \mathfrak{O}(r^2)
\,,
\\
\widehat\Gamma_A{}^1{}_B   |_{\scri^-}=& 
-\widehat\Gamma_A{}^0{}_B
-\eta_{AB} -\Big(\frac{1}{2}\theta^{(0)+} +f_1^{(1)}\Big)r\eta_{AB} +  \mathfrak{O}(r^2)
\,,
\\
\widehat\Gamma_A{}^C{}_B  |_{\scri^-} =&  \Big(1-\frac{1}{2}\theta^{+(0)} r\Big)\mathring\Gamma_A{}^C{}_B 
 +  \mathfrak{O}(r^2)
\,.
\\
\widehat\Gamma_1{}^1{}_1 |_{\scri^-}  =& 1+ f_1^{(1)}r + \mathfrak{O}(r^2)
\,,
\\
\widehat\Gamma_1{}^1{}_0 |_{\scri^-}  =&   - (\kappa^{(0)} + f_1^{(1)})r + \mathfrak{O}(r^2)
\,,
\\
\widehat\Gamma_1{}^0{}_A |_{\scri^-}  =&
\mathfrak{O}(r^2)
\,,
\\
\widehat\Gamma_1{}^1{}_A|_{\scri^-}  =&
-\widehat\Gamma_1{}^0{}_A
\,,
\\
(\widehat\Gamma_1{}^A{}_B)_{\mathrm{tf}} |_{\scri^-}  =&
0
\,,
\end{align*}
and Schouten tensor
\begin{align*}
\widehat L_{ 1 1}|_{\scri^-} 
=&  \mathfrak{O}(r^2) 
\,,
\\
\widehat L_{ 1 A}|_{\scri^-} =& 
 \mathfrak{O}(r^2)
\,,
\\
\widehat L_{ A1}|_{\scri^-} =& \mathfrak{O}(r^2)
\,,
\\
\widehat L_{ AB}|_{\scri^-} 
=&
\frac{1}{4}(\theta^{-(2)}  -3\theta^{+(0)}-4f_1^{(1)}) r\eta_{AB}
+ \mathfrak{O}(r^2)
\,,
\\
\widehat L_{ 10}|_{\scri^-} =& -( \kappa^{(0)}-\theta^{+(0)})r  + \mathfrak{O}(r^2)
\,,
\\
\widehat L_{ A 0}|_{\scri^-} 
=&  \mathfrak{O}(r^2)
\,.
\end{align*}
We further note that \cite{ttp_i0}
\begin{equation*}
(\widehat\Gamma_A{}^1{}_B  + \widehat\Gamma_A{}^0{}_B)_{\mathrm{tf}}   |_{\scri^-}=0
\end{equation*}

For the computation of the rescaled Weyl tensor 
it is in fact useful to have the coordinate components of the Schouten tensor associated to the Levi-Civita connection as well
(which we even need to some  higher orders),
\begin{align*}
 L_{rr} |_{\scri^-}
=& 
-\frac{1}{r}\theta^{+(0)}+ \mathfrak{O}(1)
\,,
\\ 
 L_{r\mathring A} |_{\scri^-}
 =& - \frac{1}{2}\mcD_A \theta^{+(1)}r
+ \mathfrak{O}(r^2)
\,,
\\
 L_{\mathring A\mathring B} |_{\scri^-}=& \frac{1}{2}s_{\mathring A\mathring B}+ \mathfrak{O}(r^2)
\,,
\\
L_r{}^r |_{\scri^-}
 =&-\frac{1}{2}  +\frac{1}{4}\Big( \kappa^{(0)}\theta^{-(2)}  +4\nu_{\tau}^{(0)} \theta^{+(0)} -\frac{3}{2}(\theta^{+(0)})^2 +\kappa^{(0)}\theta^{+(0)} \Big)r ^2 
\\
&
+\frac{1}{4}( \theta^{-(3)}  - 2\theta^{+(1)})r^2 
+\frac{1}{4}\Delta_s (\theta^{+(1)}-\kappa^{(1)})r^2
+ \mathfrak{O}(r^3)
\,,
\\
 L_{\mathring  A}{}^r|_{\scri^-} 
=&
\frac{1}{2}\mcD^{\mathring B}\Xi^{(3)}_{\mathring A\mathring B} r^3+ \frac{1}{4}\mcD_{\mathring A}(\theta^{-(3)}-\theta^{+(1)}) r^3
+ \mathfrak{O}(r^4)
\,.
\end{align*}
For the rescaled Weyl tensor we then find
\begin{align*}
W_{r\mathring Ar\mathring B}|_{\scri^-} 
 =&
-\frac{1}{2r}\Big(\Xi^{(3)}_{\mathring A\mathring B}-(\mcD_{\mathring A}\mcD_{ \mathring B}(\theta^{+(1)}-\kappa^{(1)}))_{\mathrm{tf}}\Big)
+ \mathfrak{O}(1)
\,,
\\
W_{r\mathring Ar}{}^r|_{\scri^-}
=& \frac{1}{2}\Big[ L^{(3)}_{\mathring A}{}^{r}
- \mcD_{\mathring A} L^{(2)}_{r}{}^{r}   
-\frac{1}{2}\xi^{(2)}_{\mathring A} 
  +  L^{(1)}_{r\mathring A}
\Big]
+ \mathfrak{O}(r)
\\
=&  -\frac{1}{2}\mcD^{\mathring B}W_{r\mathring Ar\mathring B}^{(-1)}
+ \mathfrak{O}(r)
\;,
\\
W_{\mathring A\mathring Br}{}^r|_{\scri^-} =& 
 \frac{1}{2}\mcD_{[\mathring A}\mcD^{\mathring C}\Xi^{(3)}_{\mathring B]\mathring C} r+ \mathfrak{O}(r^2)
\\
=&
-\mcD_{[\mathring A}\mcD^{\mathring C}W_{\mathring B] r\mathring Cr}^{(-1)} r+ \mathfrak{O}(r^2)
\;.
\end{align*}
We further obtain the ODE,
\begin{align*}
 \Big(\partial_r+\frac{3}{2} \theta^{+(0)} + \mathfrak{O}(r)\Big)W_r{}^r{}_r{}^r|_{\scri^-} 
=&
\frac{1}{2} \mcD^{\mathring A} \mcD^{\mathring B}W_{r\mathring Ar\mathring B}^{(-1)}
+ \mathfrak{O}(r)
\,,
\end{align*}
which can be integrated,
\begin{align*}
W_r{}^r{}_r{}^r|_{\scri^-} 
=&
-2m + \Big(\frac{1}{2} \mcD^{\mathring A} \mcD^{\mathring B}W_{r\mathring Ar\mathring B}^{(-1)} + 3m\theta^{+(0)}\Big)r
+\mathfrak{O}(r^2)
\,.
\end{align*}
As in \cite{ttp_i0}, the ODE
\begin{align*}
 \Big(\partial_{r}   -\frac{2}{r} + \mathfrak{O}(1)\Big) W_{\mathring A}{}^r{}_{r}{}^{r}|_{\scri^-}
=& 
  \frac{1}{2} r \mcD^{\mathring B}W^{(-1)}_{r\mathring Ar\mathring B}
+\frac{1}{2}  r\mcD_{\mathring A} W^{(1)}_{r}{}^r{}_{r}{}^r
+r W^{(0)}_{r\mathring Ar}{}^{r}  
-\frac{1}{2} r \mcD^{\mathring B}W^{(1)}_{\mathring A\mathring Br}{}^{r}
+ \mathfrak{O}(r^2)
\\
=& 
\frac{1}{4}  r\mcD_{\mathring A} \mcD^{\mathring B} \mcD^{\mathring C}W_{r\mathring Br\mathring C}^{(-1)} 
-\frac{1}{8} r \mcD^{\mathring B} \mcD_{\mathring A}\mcD^{\mathring C}\Xi^{(3)}_{\mathring B\mathring C}
+\frac{1}{8} r \Delta_s\mcD^{\mathring B}\Xi^{(3)}_{\mathring A\mathring B}
+ \mathfrak{O}(r^2)
\\
=& 
\frac{1}{2}  r\mcD_{\mathring A} \mcD^{\mathring B} \mcD^{\mathring C}W_{r\mathring Br\mathring C}^{(-1)} 
-\frac{1}{4} r (\Delta_s-1)\mcD^{\mathring B}W^{(-1)}_{r\mathring Ar\mathring B}
+ \mathfrak{O}(r^2)
\,,
\end{align*}
 leads to the \emph{no-logs condition}
\begin{equation*}
W^{(-1)}_{r\mathring Ar\mathring B}=0
\,.
\end{equation*}
If this condition is satisfied, the solution takes the form
\begin{align*}
 W_{\mathring A}{}^r{}_{r}{}^{r}|_{\scri^-}
=& 
L_{\mathring A} r^2 +  \mathfrak{O}(r^3)
\,.
\end{align*}
For the $( W_{\mathring A}{}^r{}_{\mathring B}{}^{r} )_{\mathrm{tf}} $-constraint we need to determine higher-order expansion coefficients.
For $\nabla_rW^r{}_{rr\mathring A}$ and $\nabla_rW^r{}_{r\mathring A\mathring B} $
 it is more efficient to use
the Bianchi equation (cf.\ (4.33) and (4.34) in \cite{ttp1}) and to regard directly  the radiation field $W_{r\mathring A r\mathring B}$ rather
than $\Xi_{\mathring A\mathring B}$ as the free data on $\scri^-$ (observe that we already expressed the components computed above in terms
of the radiation field),
\begin{align*}
\nabla_rW^r{}_{rr\mathring A} |_{\scri^-}= g^{\mathring C\mathring D}\nabla_{\mathring C}W_{r\mathring A r\mathring D}
\,,
\\
\nabla_rW^r{}_{r\mathring A\mathring B} |_{\scri^-}= g^{\mathring C\mathring D}\nabla_{\mathring D}W_{r\mathring C \mathring A\mathring B}
\,.
\end{align*}
We set
\begin{equation*}
w^{(0)}_{\mathring A}:= \mcD^{\mathring B}W^{(0)}_{r \mathring A r \mathring B}
\,.
\end{equation*}
Then we obtain
\begin{align*}
\Big(\partial_r+\frac{2}{r}+\mathfrak{O}(1)\Big)W^r{}_{rr\mathring A} |_{\scri^-}
=&w^{(0)}_{\mathring A} + \mathfrak{O}(r)
\\
\Longrightarrow \quad W^r{}_{rr\mathring A}|_{\scri^-}=&\frac{1}{3}w^{(0)}_{\mathring A} r+ \mathfrak{O}(r^2)
\,,
\end{align*}
and
\begin{align*}
\Big(\partial_r+ \mathfrak{O}(1)\Big)W_{\mathring A\mathring Br} {}^r
 |_{\scri^-}
=-2\mcD_{[\mathring A}W^{(1)}_{\mathring B]rr}{}^r r+  \mathfrak{O}(r^2)
=-\frac{2}{3}\mcD_{[\mathring A}w^{(0)}_{\mathring B]} r+  \mathfrak{O}(r^2)
\\
\Longrightarrow \quad W_{\mathring A\mathring Br} {}^r|_{\scri^-}=
-\frac{1}{3}\mcD_{[\mathring A}w^{(0)}_{\mathring B]}r^2+   \mathfrak{O}(r^3)
\,.
\end{align*}
Moreover, we find
%
%
\begin{align*}
\Big(\partial_r+\frac{3}{2}\theta^{+(0)} +\frac{3}{2}\theta^{+(1)}r+ \mathfrak{O}(r^2)\Big)W_r{}^r{}_r{}^r |_{\scri^-} 
 =\mcD^{\mathring C}W^{(1)r}{}_{rr\mathring C}  r +  \mathfrak{O}(r^2)
 =\frac{1}{3}\mcD^{\mathring A}w^{(0)}_{\mathring A}  r +  \mathfrak{O}(r^2)
\\
\Longrightarrow \quad W_r{}^r{}_r{}^r |_{\scri^-}=-2m + 3m\theta^{+(0)} r + \frac{1}{2}\Big(\frac{1}{3}\mcD^{\mathring A}w^{(0)}_{\mathring A}  -\frac{9}{2}m(\theta^{+(0)})^2+ 3m\theta^{+(1)}\Big)r^2+  \mathfrak{O}(r^3)
\,,
\end{align*}
as well as
\begin{align*}
 &\hspace{-3em}\Big(\partial_{r} -\frac{2}{r} +\kappa^{(0)}+\frac{1}{2}\theta^{+(0)}+  \mathfrak{O}(r) \Big) W_{\mathring A}{}^r{}_{r}{}^{r}|_{\scri^-}
\\
=& 
  \frac{1}{2} r^2\mcD^{\mathring B}W^{(0)}_{r\mathring Ar\mathring B}
+\frac{1}{2} r^2 \mcD_{\mathring A}W^{(2)}_{r}{}^r{}_{r}{}^r+\frac{3}{2}m\xi^{(2)}_{\mathring A}r^2
+r^2 W^{(1)}_{r\mathring Ar}{}^{r}  
-\frac{1}{2}  \mcD^{\mathring B}W^{(2)}_{\mathring A\mathring Br}{}^{r}r^2+  \mathfrak{O}(r^3)
\\
=& 
\frac{1}{6}\mcD_{\mathring A}  \mcD^{\mathring B}w^{(0)}_{\mathring B}r^2
-\frac{1}{12} (\Delta_s-3)w^{(0)}_{\mathring A}r^2
+\frac{3}{2}m\mcD_{ \mathring A}\Big(\theta^{+(1)}-\kappa^{(1)}\Big)r^2
+  \mathfrak{O}(r^3)
\\
\Longrightarrow \quad
W_{\mathring A}{}^r{}_{r}{}^{r}|_{\scri^-}=&L_{\mathring A} r^2 + 
\Big[\frac{1}{6}\mcD_{\mathring A}  \mcD^{\mathring B}w^{(0)}_{\mathring B}
-\frac{1}{12} (\Delta_s-3)w^{(0)}_{\mathring A}
+\frac{3}{2}m\mcD_{ \mathring A}\Big(\theta^{+(1)}-\kappa^{(1)}\Big)
\\
&
-L_{\mathring A}\Big(\kappa^{(0)}+\frac{1}{2}\theta^{+(0)}\Big)\Big]r^3
+  \mathfrak{O}(r^4)
\,.
\end{align*}
Finally, we consider the  constraint,
\begin{align*}
&\hspace{-1em} \Big(\partial_{r}   -\frac{4}{r}  + 2\kappa^{(0)} -\frac{1}{2}\theta^{+(0)}+ \mathfrak{O}(r)\Big) ( W_{\mathring A}{}^r{}_{\mathring B}{}^{r} )_{\mathrm{tf}} 
\\
=& 
(\mcD_{(\mathring A}L_{\mathring B)})_{\mathrm{tf}}r^2+  \Big( \frac{1}{6}\mcD_{\mathring A}\mcD_{\mathring B}  \mcD^{\mathring C}w^{(0)}_{\mathring C}
-\frac{1}{12}\mcD_{(\mathring A} (\Delta_s-1)w^{(0)}_{\mathring B)}
-\Big(\kappa^{(0)}+\frac{1}{2}\theta^{+(0)}\Big)\mcD_{(\mathring A}L_{\mathring B)}
\Big)_{\mathrm{tf}}  r^3
+  \mathfrak{O}(r^4)
\;.
\end{align*}
This equation comes along with the following no-logs condition, 
\begin{equation*}
\Big(2\mcD_{\mathring A}\mcD_{\mathring B}  \mcD^{\mathring C}w^{(0)}_{\mathring C}
-\mcD_{(\mathring A} (\Delta_s-1)w^{(0)}_{\mathring B)}
+12(\kappa^{(0)}-\theta^{+(0)})\mcD_{(\mathring A}L_{\mathring B)}\Big)_{\mathrm{tf}}=0
\,,
\end{equation*}
which, on the round sphere, is equivalent to its divergence,
\begin{equation}
\mcD_{\mathring A}  (\Delta_s + 2)\mcD^{\mathring B}w^{(0)}_{\mathring B}
-\frac{1}{2}(\Delta_s+1) (\Delta_s-1)w^{(0)}_{\mathring A}
=
-6(\kappa^{(0)}-\theta^{+(0)})(\Delta_s+1)L_{\mathring A}
\,.
\label{no-logs_fin_constr}
\end{equation}
If this condition is satisfied, we have
\begin{equation*}
 ( W_{\mathring A}{}^r{}_{\mathring B}{}^{r} )_{\mathrm{tf}} =-(\mcD_{(\mathring A}L_{\mathring B)})_{\mathrm{tf}}r^3 + \mathring c^{(2,0)}_{\mathring A\mathring B} r^4 + \mathfrak{O}(r^5)
\,.
\end{equation*}
To analyze the no-logs condition \eq{no-logs_fin_constr}
we make a Hodge decomposition of $L_{\mathring A}$ and $W^{(0)}_{r\mathring Ar \mathring B}$,
\begin{align*}
L_{\mathring A}=&\mcD_{\mathring A} \ul L + \epsilon_{\mathring A}{}^{\mathring B}\mcD_{\mathring B}\ol L
\,,
\\
W^{(0)}_{r\mathring Ar \mathring B}=&(\mcD_{\mathring A}\mcD_{\mathring B} \ul w^{(0)})_{\mathrm{tf}}+  \epsilon_{(\mathring A}{}^{\mathring C}\mcD_{\mathring B)}\mcD_{\mathring C}\ol w^{(0)}
\,.
\end{align*}
Then the no-logs-condition becomes
\begin{align*}
 (\Delta_s + 2)\Delta_s\Big( (\Delta_s+2)\Delta_s\ul w^{(0)}
+24(\kappa^{(0)}-\theta^{+(0)})\ul L\Big)=&0
\,,
\\
(\Delta_s+2)\Delta_s\Big( (\Delta_s+2)\Delta_s\ol w^{(0)}
+24(\kappa^{(0)}-\theta^{+(0)})\ol L\Big)
=&0
\,.
\end{align*}
In order to fulfill these no-logs conditions one can prescribe an arbitrary expansion coefficient $W^{(0)}_{r \mathring A r \mathring B}$
of the radiation field (equivalently $V^{+(2)}_{AB}$), which then determines the tensor $L_{\mathring A}$ up to the addition of conformal Killing vectors on $\mathbb{S}^2$.
In particular whenever $L_{\mathring A}$ is a conformal Killing vector, $W^{(0)}_{r \mathring A r \mathring B}$ needs to vanish.
However, even if the no-logs condition is satisfied, a non-trivial $W^{(0)}_{r \mathring A r \mathring B}$  may produce log-terms in higher orders.
This will  be analyzed further in Section~\ref{sec_spec_gauge} below.

On the other hand, if the radiation field, or rather, at this order, the expansion coefficient  $V^{+(2)}_{AB}$, vanishes,
$L_A$ needs to be a conformal Killing vector whenever $\kappa^{(0)}-\theta^{+(0)}\ne 0$. This contrasts a gauge where $\kappa^{(0)}-\theta^{+(0)}= 0$
where it has been shown in \cite{ttp_i0} that no logarithmic  terms arise for any choice of $L_A$ whatsoever.

By way of summary,
if all no-logs condition are satisfied, here expressed in terms of frame coefficients,
%
%
\begin{align}
&V^{+(0)}_{AB}=0
\,,\quad  V^{+(1)}_{AB}=0
\,,
\label{no_logs_pre}
\\
&\mcD_{ A}  (\Delta_s + 2)\mcD^{ B}\mcD^CV^{+(2)}_{ BC}
-\frac{1}{2}(\Delta_s+1) (\Delta_s-1)\mcD^BV^{+(2)}_{ AB}
=
-3(\kappa^{(0)}-\theta^{+(0)})(\Delta_s+1)L_{ A}
\,.
\label{first_no-logs_cond}
\end{align}
%
The frame coefficients of the  rescaled Weyl tensor have the following expansions at $I^-$,
\begin{align*}
W_{0101}|_{\scri^-}=&-2m + 3m\theta^{+(0)} r +\Big(\frac{1}{3} \mcD^A\mcD^BV^{+(2)}_{AB}-\frac{9}{4}m(\theta^{+(0)})^2+\frac{3}{2}m\theta^{+(1)}\Big)r^2+ \mathfrak{O}(r^3)
\,,
\\
W_{01AB}|_{\scri^-}= &\frac{2}{3} \mcD_{[A} \mcD^CV^{+(2)}_{B]C} r^2+ \mathfrak{O}(r^3)
\,,
\\
W^-_{A}|_{\scri^-} =&- \frac{2}{3} \mcD^BV^{+(2)}_{AB}r^2+  \mathfrak{O}(r^3)
\,,
\\
W^+_{A} |_{\scri^-}=&2L_A r
+\Big(
\frac{2}{3}\mcD_A \mcD^B\mcD^CV^{+(2)}_{BC}
-\frac{1}{3}(\Delta-1) \mcD^BV^{+(2)}_{AB}
\\
&
-2(\kappa^{(0)}+ \theta^{+(0)}-\nu^{(0)}_{\tau})L_A 
+6m\Big( \nu^{(2)}_{ A}  + \frac{1}{2} \mcD_{ A}(\theta^{+(1)}-\kappa^{(1)})  \Big)
\Big)r^2+ \mathfrak{O}(r^3)
\,,
\\
V^+_{AB}|_{\scri^-} =&V^{+(2)}_{AB} r^2+ \mathfrak{O}(r^3)
\,,
\\
V^-_{AB}|_{\scri^-}=& 
 -2(\mcD_{( A}L_{ B)})_{\mathrm{tf}}r+  c^{(2,0)}_{ A B} r^2
+ \mathfrak{O}(r^3)
\,.
\end{align*}

We further find that (we do not need the components of the Schouten tensor)
\begin{align*}
\partial_{\tau}\widehat\Gamma_{1}{}^0{}_1 |_{\scri^-}
 &=
   \mathfrak{O}(r^2)
\,,
\\
\partial_{\tau}\widehat\Gamma_{1}{}^1{}_1  |_{\scri^-}
 &=
(\theta^{+(0)} + f_1^{(1)})r
+   \mathfrak{O}(r^2)
\,,
\\
\partial_{\tau}\widehat\Gamma_{1}{}^0{}_A  |_{\scri^-}
 &=
  \mathfrak{O}(r^2)
\,,
\\
\partial_{\tau}\widehat\Gamma_{1}{}^1{}_A  |_{\scri^-}
 &=
 \mathfrak{O}(r^2)
\,,
\\
\partial_{\tau}\widehat\Gamma_{A}{}^0{}_1 |_{\scri^-}
 &=
 \mathfrak{O}(r^2)
\,,
\\
\partial_{\tau}\widehat\Gamma_{A}{}^1{}_1  |_{\scri^-}
 &=
 \mathfrak{O}(r^2)
\,,
\\
\partial_{\tau}(\widehat\Gamma_{A}{}^0{}_B -\widehat\Gamma_{A}{}^1{}_B ) |_{\scri^-}
 &=
\frac{1}{2}(\theta^{-(2)}  -\theta^{+(0)}-2f_1^{(1)}) r\eta_{AB}+   \mathfrak{O}(r^2)
\,,
\\
\partial_{\tau}(\widehat\Gamma_{A}{}^0{}_B + \widehat\Gamma_{A}{}^1{}_B ) |_{\scri^-}
 &=
-(  \theta^{+(0)}+f_1^{(1)}) r\eta_{AB}  +   \mathfrak{O}(r^2)
\,,
\\
(\partial_{\tau}\widehat\Gamma_{1}{}^A{}_B )_{\mathrm{tf}} |_{\scri^-}
 &=
 \mathfrak{O}(r^2)
\,,
\\
\partial_{\tau}\widehat\Gamma_{A}{}^B{}_C  |_{\scri^-}
 &=
 \frac{1}{4}\Big( \theta^{-(2)} + \theta^{+(0)}\Big)r \mathring\Gamma_A{}^B{}_C
 +   \mathfrak{O}(r^2)
\,,
\\
\partial_{\tau}e^{\tau}{}_1 |_{\scri^-}
&= - 1 +\kappa^{(0)} r +   \mathfrak{O}(r^2)
\,,
\\
\partial_{\tau}e^{r}{}_1 |_{\scri^-}
&=
  (\kappa^{(0)} + f_1^{(1)})r^2
 +   \mathfrak{O}(r^3)
\,,
\\
\partial_{\tau}e^{\mathring A}{}_1 |_{\scri^-}
&= 
 \mathfrak{O}(r^2)
\,,
\\
\partial_{\tau}e^{\tau}{}_A |_{\scri^-}
&=  \mathfrak{O}(r^2)
\,,
\\
\partial_{\tau}e^r{}_A |_{\scri^-}
&=
   \mathfrak{O}(r^3)
\,,
\\
\partial_{\tau}e^{\mathring A}{}_A |_{\scri^-}
&= 
  \frac{1}{4}\Big( \theta^{-(2)} + \theta^{+(0)}\Big)r\mathring e^{\mathring A}{}_A+   \mathfrak{O}(r^2)
\,,
\\
\partial_{\tau}^2e^{\tau}{}_1 |_{\scri^-}
&= -(\theta^{+(0)} + \kappa^{(0)} +2 f_1^{(1)})r
+   \mathfrak{O}(r^2)
\,.
\end{align*}

\subsection{Higher orders}

Recall that the functions $c_{AB}^{(n+2,n)}:=\frac{1}{(n+2)!}\partial_r^{n+2}\partial_{\tau}^n V^-_{AB}|_{I^-}$, $n\geq 0$, on $I^-$ may be regarded as part of the freely prescribable data on $I^-$ \cite{ttp_i0}.
We want to keep track in which way  these functions enter the no-logs conditions in higher orders.
In the following we will consider transverse derivatives on $\scri^-$ of order $n+1$, $n \geq0$,  where we assume that all transverse derivatives
up to and including the order $n$ are smooth at $I^-$.
We write
\begin{equation*}
f|_{\scri^-}=\mathfrak{O}_{ n-1}
\end{equation*}
if $f|_{\scri^-}$ is smooth at $I^-$ and only depends on the radiation field, $m$, $L_A$,  the gauge data and the functions 
$c_{AB}^{k+2,k}$ with $k\leq n-1$ but not on those where $k\geq n$.

From \eq{transport1}-\eq{transport22}
we immediately obtain for \underline{$k\leq n$}
%
\begin{align*}
\partial_{\tau}^{ k+1}\widehat L_{ij}|_{\scri^-} =& \mathfrak{O}_{ k}
\,,
\\
\partial_{\tau}^{ k+1}\widehat \Gamma_{i}{}^k{}_{j}|_{\scri^-} =& \mathfrak{O}_{ k-1}
\,,
\\
\partial_{\tau}^{ k+1}e^{\mu}{}_{i}|_{\scri^-} =& \mathfrak{O}_{ k-2}
\,.
\end{align*}
For the rescaled Weyl tensor we find from \cite[Equations (2.90)-(2.95)]{ttp_i0} for \underline{$k\leq n$}
\begin{align*}
\partial_{\tau}^{k+1}W_{0101} |_{\scri^-}
 =& 
 \frac{1}{2}(\widehat\Gamma^{AB}{}_{1}-\widehat\Gamma^{AB}{}_{0})\partial_{\tau}^kV^-_{AB}+\mathfrak{O}_{ k-1}
= \mathfrak{O}_{ k-1}
\,,
\\
\partial_{\tau}^{k+1}W_{01AB} |_{\scri^-}     =&
( \widehat\Gamma^{C1}{}_{[A}+ \widehat\Gamma^{C0}{}_{[A})\partial_{\tau}^k V^-_{B]C}+\mathfrak{O}_{ k-1}=\mathfrak{O}_{ k-1}
\,,
\\
\partial_{\tau}^{k+1}W^-_A|_{\scri^-}
 =&\mathfrak{O}_{ k-1}
\,,
\\
\partial_{\tau} ^{k+1} V^+_{AB}|_{\scri^-}
 =&\mathfrak{O}_{ k-1}
\,,
\\
\partial_{\tau}^{k+1}W^+_A|_{\scri^-}
 =&( -  \check\nabla^B
+3 \widehat\Gamma^{B0}{}_{0}
+2\widehat\Gamma^{B0}{}_1)\partial_{\tau}^kV^-_{AB}
+\mathfrak{O}_{ k-1}
\,,
\end{align*}
and also
\begin{equation}
 \partial_{\tau}^k V^-_{AB} |_{\scri^-}
= c^{(k+2,k)}_{AB} r^{k+2}+\Big((k+2)(\nu^{(0)}_{\tau} -  \kappa^{(0)} )-\frac{1}{2}\theta^{(0)+} \Big)c^{(k+2,k)}_{AB} r^{k+3}
+ \mathfrak{O}(r^{k+4}) +\mathfrak{O}_{ k-1}
\,,
\label{expansion_tran_V-}
\end{equation}
where
$\check\nabla$ is defined by (similarly for  tensors of higher valence),
\begin{equation}
\check\nabla_Av_B := e^{\mu}{}_A\partial_{\mu}v_B -\widehat\Gamma_A{}^C{}_Bv_C
\,.
\end{equation}
Finally, we consider the equation for $\partial_{\tau} ^{n+1} V^-_{AB}|_{\scri^-}$ \cite[Equation (2.95)]{ttp_i0}
\begin{align*}
&\hspace{-2em}\Big( \nu^{\tau}\partial_{r}  + (n+1) \partial_{\tau} e^{\tau}{}_1
-3f_1 -2 \widehat\Gamma_1{}^1{}_{0} 
  - \frac{1}{2}\widehat\Gamma_C{}^C{}_0 + \frac{1}{2}\widehat\Gamma_C{}^C{}_1 \Big)\partial_{\tau}^{n+1} V^-_{AB} |_{\scri^-}
\\
=&
-\begin{pmatrix}n+1\\2\end{pmatrix} \partial_{\tau} ^2 e^{\tau}{}_1\partial_{\tau}^n V^-_{AB} 
-(n+1)\partial_{\tau} e^{\alpha}{}_1\partial_{\alpha} \partial_{\tau}^n V^-_{AB} 
+(n+1)\partial_{\tau}  \Big(3f_1+ 
2 \widehat\Gamma_1{}^1{}_{0} 
  +\frac{1}{2} \widehat\Gamma_C{}^C{}_0 
-\frac{1}{2} \widehat\Gamma_C{}^C{}_1 \Big) \partial_{\tau}^n V^-_{AB} 
\\
&
+ \Big( (\check\nabla_{(A}
- 4\widehat\Gamma_1{}^0{}_{(A} 
- 2 f_{(A}
- \widehat\Gamma_{(A}{}^0{}_{|1|} )\partial_{\tau} ^{n+1}W^+_{B)}
\Big)_{\mathrm{tf}}+\mathfrak{O}_{ n-1}
\,,
\end{align*}
which yields
%
%
\begin{align*}
&\hspace{-2em}\Big( (r-\nu^{(0)}_{\tau}r^2 + \mathfrak{O}(r^3))\partial_{r}  -  n-3 + (n+3) \kappa^{(0)} r 
+\frac{1}{2}\theta^{(0)+} r+   \mathfrak{O}(r^2) \Big)\partial_{\tau}^{n+1} V^-_{AB} |_{\scri^-}
\\
=&
\frac{1}{2}(n+1)(n+4)(\theta^{+(0)} -\kappa^{(0)})c_{AB}^{(n+2,n)}r^{n+3}
- \Big( \mcD_{(A} \mcD_Cc_{B)}{}^{C(n+2,n)}\Big)_{\mathrm{tf}}r^{n+2}
\\
&
- \Big((n+2)(\nu^{(0)}_{\tau} -  \kappa^{(0)}) -\frac{3}{2}\theta^{(0)+} \Big)\Big( \mcD_{(A} \mcD_Cc_{B)}{}^{C(n+2,n)}\Big)_{\mathrm{tf}}r^{n+3}
+\mathfrak{O}_{ n-1}+\mathfrak{O}(r^{n+4})
\,.
\end{align*}
The no-logs condition therefore adopts the form
\begin{align*}
(\kappa^{(0)} -\theta^{+(0)} )\Big((n+1)(n+4)c_{AB}^{(n+2,n)}
+ 2( \mcD_{(A} \mcD_Cc_{B)}{}^{C(n+2,n)})_{\mathrm{tf}}\Big)
=\mathfrak{O}_{ n-1}
\,,
\end{align*}
where the right-hand side is a symmetric trace-free tensor, or, equivalently,
\begin{align*}
(\kappa^{(0)} -\theta^{+(0)} )\Big(\Delta_s + (n+2)(n+3)-1\Big) \mcD_Bc_{A}{}^{B(n+2,n)}
=\mathfrak{O}_{ n-1}
\,.
\end{align*}
If this condition is fulfilled, \eq{expansion_tran_V-} also holds for $k=n+1$.

To analyze the solvability of this equation it is useful to know in which way the expansion coefficients 
of the radiation field enter this condition. In fact, this computation has already been accomplished in \cite{ttp_i0},
and remains valid within the larger class of gauges we allow here.

It has been shown there that
\begin{align*}
&\hspace{-2em}\Big( (r-\nu^{(0)}_{\tau}r^2 + \mathfrak{O}(r^3))\partial_{r}  -  n-3 + (n+3) \kappa^{(0)} r 
+\frac{1}{2}\theta^{(0)+} r+   \mathfrak{O}(r^2) \Big)\partial_{\tau}^{n+1} V^-_{AB} |_{\scri^-}
\\
=&
\frac{1}{2}(n+1)(n+4)(\theta^{+(0)} -\kappa^{(0)})c_{AB}^{(n+2,n)}r^{n+3}
- \Big( \mcD_{(A} \mcD_Cc_{B)}{}^{C(n+2,n)}\Big)_{\mathrm{tf}}r^{n+2}
\\
&
- \Big((n+2)(\nu^{(0)}_{\tau} -  \kappa^{(0)}) -\frac{3}{2}\theta^{(0)+} \Big)\Big( \mcD_{(A} \mcD_Cc_{B)}{}^{C(n+2,n)}\Big)_{\mathrm{tf}}r^{n+3}
\\
&
+\frac{(-1)^{n}}{2^{n}(n+4)!}  \prod_{\ell =2}^{n+2}\Big(\mcD_{(A} \Big(\Delta_s+\ell (\ell+1)-1\Big)\Big(( \Delta_s-1)v^{(n+5)}_{ B)}
-2 \mcD_{ B)}\mcD^Cv^{(n+5)}_{ C}\Big)\Big)_{\mathrm{tf}}r^{n+3}
\\
&
+ \widehat{\mathfrak{O}}_{ n-1}
+\mathfrak{O}(r^{n+4})
\,,
\end{align*}
where we write
\begin{equation*}
f|_{\scri^-}=\widehat{\mathfrak{O}}_{ n-1}
\end{equation*}
if $f|_{\scri^-}$ is smooth at $I^-$ and only depends on  $m$, $L_A$,  the gauge data,
$c_{AB}^{k+2,k}$, and the expansion coefficients of the radiation field
$V^{+(k+3)}_{AB}$
 with $k\leq n-1$ but not on those where $k\geq n$ (i.e.\ as compared with $\mathfrak{O}$  only  expansion coefficients  of the radiation field up to a certain order are allowed).

A more sophisticated version of the no-logs conditions therefore takes the form 
\begin{align*}
&
\frac{(-1)^{n}}{2^{n-1}(n+4)!}  \prod_{\ell =2}^{n+2}\Big(\mcD_{(A} \Big(\Delta_s+\ell (\ell+1)-1\Big)\Big(( \Delta_s-1)v^{(n+5)}_{ B)}
-2 \mcD_{ B)}\mcD^Cv^{(n+5)}_{ C}\Big)\Big)_{\mathrm{tf}}
\\
&
+(\kappa^{(0)} -\theta^{+(0)} )\Big((n+1)(n+4)c_{AB}^{(n+2,n)}
+ 2( \mcD_{(A} \mcD_Cc_{B)}{}^{C(n+2,n)})_{\mathrm{tf}}\Big)
=
 \widehat{\mathfrak{O}}_{ n-1}
\,,
\end{align*}
where the right-hand side is a symmetric trace-free tensor, or, equivalently
\begin{align*}
&
\frac{(-1)^{n}}{2^{n}(n+4)!}  \prod_{\ell =1}^{n+2}\Big(\Delta_s+\ell (\ell+1)-1\Big)\Big(( \Delta_s-1)v^{(n+5)}_{ A}
-2 \mcD_{ A}\mcD^Cv^{(n+5)}_{ C}\Big)
\\
&
+(\kappa^{(0)} -\theta^{+(0)} )\Big(\Delta_s + (n+2)(n+3)-1\Big) \mcD_Bc_{A}{}^{B(n+2,n)}
=
 \widehat{\mathfrak{O}}_{ n-1}
\,.
\end{align*}
Divergence and curl read
\begin{align*}
\prod_{\ell =0}^{n+2}\Big(\Delta_s+\ell (\ell+1)\Big) \mcD^Av^{(n+5)}_{ A}
-(\kappa^{(0)} -\theta^{+(0)} )\Big(\Delta_s + (n+2)(n+3)\Big)\mcD^A \mcD^Bc_{AB}{}^{(n+2,n)}
=
 \widehat{\mathfrak{O}}_{ n-1}
\,,
\\
 \prod_{\ell =0}^{n+2}\Big(\Delta_s+\ell (\ell+1)\Big)\mcD_{[A}v^{(n+5)}_{ B]}
+(\kappa^{(0)} -\theta^{+(0)} )\Big(\Delta_s + (n+2)(n+3)\Big) \mcD_{[A}\mcD^Cc_{B]C}{}^{(n+2,n)}
=
 \widehat{\mathfrak{O}}_{ n-1}
\,.
\end{align*}
One would like to read the no-logs conditions as equations on the expansion coefficients of the radiation field and/or the $c_{AB}^{(n+2,n)}$.
However, the equation can only be solved on $\mathbb{S}^2$ supposing that the right-hand side does not contain
$\ell=n+2$-spherical harmonics in its spherical decomposition.
As in \cite{ttp_i0}, though,  one should expect that e.g. a non-trivial  expansion coefficient of the radiation field will eventually produce
$\ell=n+2$-spherical harmonics on the right-hand side for some $n$. So, as in \cite{ttp_i0}, one should expect severe restrictions on the
data by the no-logs condition, which is to be analyzed elsewhere.
In order to get some intuition we will consider in the following section the next order explicitly for a more specific choice of the gauge data.

\subsection{Special gauge}
\label{sec_spec_gauge}

\subsubsection{Solution of the constraints}

Let us  choose gauge data as follows,
\begin{align}
\Theta^{(1)} = 2 re^{-\kappa^{(0)} r}
\,,\quad
\nu^{\tau} = r
\,,
\quad
\nu_{\mathring A} =  0
\,,
\label{spec_gauge1}
\\
\kappa= -\frac{2}{r}+ \kappa^{(0)}
\,,\quad
\theta^- =0
\,,
\quad
g_{\mathring A\mathring B} |_I = s_{\mathring A\mathring B}
\\
f_1|_{\scri^-} = 1
\,,
\quad
f_{ A}|_{\scri^-} =0
\,.
\label{spec_gauge3}
\end{align}
(As in \cite{ttp_i0} it would suffice for the following computation if the gauge data satisfy these relations
only up to each order at $I^-$, in which case one would have to add $\mathfrak{O}(r^{\infty})$-terms everywhere.)
Then
\begin{equation*}
e_0|_{\scri^-}=\partial_{\tau}
\,,
\quad
e_1|_{\scri^-}= \partial_{\tau} +r\partial_r
\,,
\quad
e_A|_{\scri^-} =\mathring   e ^{\mathring A}{}_A\partial_{\mathring A} 
\,,
\end{equation*}
and, globally,
\begin{align}
\Theta=re^{-\kappa^{(0)} r}(1-\tau^2)\,, 
\quad
b_0=-2r\tau e^{-\kappa^{(0)} r} \,, \quad
b_1=2re^{-\kappa^{(0)} r}\,, \quad
b_A=0
\,.
\label{spec_gauge_global_terms}
\end{align}
As solutions of the constraint equations \cite[Appendix~A]{ttp_i0} we obtain
\begin{align*}
\theta^+= &0
\,,
\quad
\Omega=1
\,,
\quad \xi_{\mathring A} = 0
\,,
\\
L_{rr}|_{\scri^-} =& 0
\,,
\quad
L_{r\mathring A}|_{\scri^-} =0
\,,
\quad
g^{\mathring A\mathring B} L_{\mathring A\mathring B}|_{\scri^-} =1
\,,
\\
L_{r}{}^r|_{\scri^-} =&  - \frac{1}{2}
\,,
\quad
R|_{\scri^-}= 0
\,,
\\
 (L_{\mathring A\mathring B})_{\mathrm{tf}}|_{\scri^-} =& -\frac{1}{2}\Big(\partial_r-\frac{2}{r}+\kappa^{(0)}\Big)\Xi_{\mathring A\mathring B}
\,,
\\
L_{\mathring A}{}^r|_{\scri^-} =&  \frac{1}{2}\mcD^B\Xi_{\mathring A\mathring B}
\,,
\end{align*}

From  \cite[Section~2.7 \& 2.8]{ttp_i0}
we compute 
\begin{align*}
\widehat L_{ 1 1}|_{\scri^-} =&\widehat L_{ 1 A}|_{\scri^-} 
=0
\,,
\\
\widehat L_{ A1}|_{\scri^-} =&\frac{1}{2r}\mcD^B\Xi_{ A B}
\,,
\\
\widehat L_{ AB}|_{\scri^-} 
=&
 -\frac{1}{2}\Big(\partial_r-\frac{1}{r}+\kappa^{(0)}\Big)\Xi_{ A B} 
\,,
\\
\widehat L_{ 10}|_{\scri^-} =& -\kappa^{(0)}r
\,,
\\
\widehat L_{ A 0}|_{\scri^-} 
=&\frac{1}{2r}\mcD^B\Xi_{ A B}
\,,
\end{align*}
%
and
\begin{align*}
 \widehat\Gamma_A{}^1{}_1   |_{\scri^-}=& \widehat\Gamma_A{}^1{}_0   |_{\scri^-}=\widehat\Gamma_1{}^0{}_A |_{\scri^-}  =\widehat\Gamma_1{}^1{}_A|_{\scri^-}  =(\widehat\Gamma_1{}^A{}_B)_{\mathrm{tf}} |_{\scri^-}  =0
\,,
\\
 \widehat\Gamma_A{}^B{}_0|_{\scri^-}  =& 
 \frac{1}{2r} \Xi_{ A}{}^{ B} 
\,,
\\
\widehat\Gamma_A{}^B{}_1   |_{\scri^-}=& 
\delta^{B}{}_A+ \frac{1}{2r} \Xi_{ A}{}^{ B} 
\,,
\\
\widehat\Gamma_A{}^C{}_B  |_{\scri^-} =&    \mathring\Gamma_A{}^C{}_B 
\,,
\\
\widehat\Gamma_1{}^1{}_1 |_{\scri^-}  =& 1
\,,
\\
\widehat\Gamma_1{}^1{}_0 |_{\scri^-}  =&   -\kappa ^{(0)}r 
\,.
\end{align*}
For the rescaled Weyl tensor we find from \cite[Equations (2.95)-(2.99)]{ttp_i0}
\begin{align*}
(r\partial_r +1-\kappa ^{(0)}r )W^-_{A}  |_{\scri^-}  
 =&- 2 \mcD^BV^+_{AB}
\,,
\\
r\partial_r W_{0101} |_{\scri^-}  
 =& 
-\mcD^AW^{-}_A
+ \frac{1}{r} \Xi^{A B} V^+_{AB}
\,,
\\
r\partial_r W_{01AB}   |_{\scri^-}   =&
-2\mcD_{[A} W^{-}_{B]}
+ \frac{2}{r} \Xi_{[A}{} ^C V^+_{B]C}
\,,
\\
(r\partial_r -1+ \kappa ^{(0)}r )W^+_{A}  |_{\scri^-}    =&
 \mcD_A W_{0101}+  \mcD^B W_{01AB}
-  \frac{2}{r} \Xi_{ AB} W^{-B}
\,,
\\
(r\partial_{r}-2+2\kappa^{(0)}r) V^-_{AB}  |_{\scri^-}  
=& \Big(  \frac{3}{2r} \Xi_{ AB} W_{0101}
+  \frac{3}{2r} \Xi_{C (A}W_{01 B)}{}^C
+\mcD_{(A}W^+_{B)}
\Big)_{\mathrm{tf}}
\,.
\end{align*}
Recall that in our current setting
\begin{equation*}
M=-m=\mathrm{const.}\,, \quad N=0\,, \quad V^+_{ A B}=V^{+(2)}_{AB}r^2+ V^{+(3)}_{AB} r^3
+ \mathfrak{O}(r^4)
\,.
\end{equation*}
Moreover, we have
\begin{equation*}
V^+_{AB} |_{\scri^-}   = 
\frac{1}{2}r^2 W_{r Ar B}
=-\frac{1}{8}e^{\kappa^{(0)} r}\partial_r\Big(\partial_r-\frac{2}{r} + \kappa^{(0)}\Big)\Xi_{ A B}
\quad \Longrightarrow \quad 
\Xi^{(4)}_{ A B}=-\frac{4}{3}V^{+(2)}_{AB}
\,.
\end{equation*}
It is convenient to set
\begin{equation*}
V^+_{A} := \mcD^B V^+_{AB}
\,.
\end{equation*}
Solving the ODEs yields the following expansions at $I^-$
\begin{align*}
W^-_{A}  |_{\scri^-} 
 =&- \frac{2}{3} V^{+(2)}_{A}r^2- \frac{1}{2}\Big( V^{+(3)}_{A}+\frac{1}{3}\kappa^{(0)}V^{+(2)}_{A}\Big)r^3
+ \mathfrak{O}(r^4)
\,,
\\
 W_{0101} |_{\scri^-} 
 =& -2m+
 \frac{1}{3} \mcD^AV^{+(2)}_{A}r^2+\frac{1}{6}\Big(\mcD^A V^{+(3)}_{A}+\frac{1}{3}\kappa^{(0)}\mcD^AV^{+(2)}_{A}\Big)r^3
+ \mathfrak{O}(r^4)
\,,
\\
W_{01AB}   |_{\scri^-}  =&
 \frac{2}{3}\mcD_{[A} V^{+(2)}_{B]}r^2+\frac{1}{3}\Big(\mcD_{[A} V^{+(3)}_{B]}+\frac{1}{3}\kappa^{(0)}\mcD_{[A}V^{+(2)}_{B]}\Big)r^3
+ \mathfrak{O}(r^4)
\,,
\\
W^+_{A}  |_{\scri^-}   =& 2L_{ A} r + \Big(
 \frac{2}{3} \mcD_A \mcD^BV^{+(2)}_{B}-  \frac{1}{3}(\Delta_s-1) V^{+(2)}_{A}-2\kappa^{(0)}L_A
\Big)r^2
\\
&
+\Big[
\frac{1}{6} \mcD_A\Big(\mcD^B V^{+(3)}_{B}-\frac{5}{3}\kappa^{(0)}\mcD^BV^{+(2)}_{B}\Big)
-\frac{1}{12}( \Delta_s-1) \Big(V^{+(3)}_{A}-\frac{5}{3}\kappa^{(0)} V^{+(2)}_{A}\Big)
\\
&
+ (\kappa^{(0)})^2L_A
\Big]r^3
+\mathfrak{O}(r^4)
\,,
\\
V^-_{A} |_{\scri^-}   =& -(\Delta_s+ 1)L_{ A} r + c^{(2,0)}_Ar^2
+ \Big[
\frac{1}{12} (\Delta_s+ 1)\mcD_A\Big(\mcD^B V^{+(3)}_{B}-\frac{5}{3}\kappa^{(0)} \mcD^B V^{+(2)}_{B}\Big)
\\
&
-\frac{1}{24}(\Delta_s+ 1)( \Delta_s-1) \Big(V^{+(3)}_{A}-\frac{5}{3}\kappa^{(0)} V^{+(2)}_{A}\Big)
+ 4mV^{+(2)}_A r^3
\\
&
+\frac{1}{2} (\kappa^{(0)})^2(\Delta_s+ 1)L_A
-2\kappa^{(0)}c^{(2,0)}_A
\Big]r^3 
+ \mathfrak{O}(r^4)
\,,
\end{align*}
supposing that the no-logs condition \eq{first_no-logs_cond} is fulfilled, i.e.
\begin{align}
2(\Delta_s+ 1) \mcD_A \mcD^BV^{+(2)}_{B}-(\Delta_s+ 1) (\Delta_s-1) V^{+(2)}_{A}
+6\kappa^{(0)}(\Delta_s+ 1)L_{ A} =0
\,.
\label{main_no-logs1}
\end{align}

\subsubsection{Transverse derivatives on $\scri^-$}

The restrictions of all transverse derivatives of the Schouten tensor to $\scri^-$ are  smooth at $I^-$, its explicit form is not needed. For connection and frame coefficients  we find
\begin{align*}
\partial_{\tau}\widehat\Gamma_{1}{}^0{}_1 |_{\scri^-}
 &=
- (\kappa ^{(0)})^2r ^2
\,,
\\
\partial_{\tau}\widehat\Gamma_{1}{}^1{}_1 |_{\scri^-}
 &=
0
\,,
\\
\partial_{\tau}\widehat\Gamma_{1}{}^0{}_A |_{\scri^-}
 &=
0
\,,
\\
\partial_{\tau}\widehat\Gamma_{1}{}^1{}_A |_{\scri^-}
 &=0
\,,
\\
\partial_{\tau}(\widehat\Gamma_{A}{}^0{}_1-\widehat\Gamma_{A}{}^1{}_1 )|_{\scri^-}
 &=
0
\,,
\\
\partial_{\tau}(\widehat\Gamma_{A}{}^0{}_1+\widehat\Gamma_{A}{}^1{}_1 )|_{\scri^-}
 &=\frac{1}{r}\mcD^B\Xi_{ A B}
\,,
\\
\partial_{\tau}(\widehat\Gamma_{A}{}^0{}_B -\widehat\Gamma_{A}{}^1{}_B )|_{\scri^-}
 &=
 -\frac{1}{2}\Big(\partial_r+\kappa^{(0)}\Big)\Xi_{ A B}   
-   \frac{1}{4r^2}|\Xi|^2\eta_{AB}
\,,
\\
\partial_{\tau}(\widehat\Gamma_{A}{}^0{}_B + \widehat\Gamma_{A}{}^1{}_B )|_{\scri^-}
 &=
 -\frac{1}{2}\Big(\partial_r-\frac{2}{r}+\kappa^{(0)}\Big)\Xi_{ A B}  
\,,
\\
(\partial_{\tau}\widehat\Gamma_{1}{}^A{}_B )_{\mathrm{tf}}|_{\scri^-}
 &=
0
\,,
\\
\partial_{\tau}\widehat\Gamma_{A}{}^B{}_C |_{\scri^-}
 &=
 - \frac{1}{2r}  \mathring \Gamma_D{}^B{}_C\Xi_{ A}{}^{ D} 
+ \frac{1}{2r}\delta^B{}_{C}\mcD^D\Xi_{ A D}
\,,
\\
\partial_{\tau}e^{\tau}{}_1|_{\scri^-}&= -1+\kappa ^{(0)}r 
\,,
\\
\partial_{\tau}e^{r}{}_1|_{\scri^-}&=
\kappa ^{(0)}r ^2
\,,
\\
\partial_{\tau}e^{\mathring A}{}_1|_{\scri^-}&= 
0
\,,
\\
\partial_{\tau}e^{\tau}{}_A|_{\scri^-}&= 
0
\,,
\\
\partial_{\tau}e^r{}_A|_{\scri^-}&=
0
\,,
\\
\partial_{\tau}e^{\mathring A}{}_A|_{\scri^-}&= 
 - \frac{1}{2r} \Xi_{ A}{}^{ B}  \mathring e^{\mathring A}{}_B
\,.
\end{align*}

We also need to determine,
\begin{align*}
\partial_{\tau}W^+_A|_{\scri^-}
 =& -  V^-_{A}
 -\frac{1}{2}\mcD_AW_{0101}
 -\frac{1}{2}\mcD^B W_{01AB}
 -  W^+_A
+ \frac{1}{r} \Xi_{ A}{}^{ B} W^{-B}
\\
=&
 (\Delta_s- 1)L_{ A} r - c^{(2,0)}_Ar^2
 - \Big( \mcD_A \mcD^BV^{+(2)}_{B}-  \frac{1}{2}(\Delta_s-1) V^{+(2)}_{A}-2\kappa^{(0)}L_A
\Big)r^2
\\
&
- \Big[
\frac{1}{12} (\Delta_s+ 5)\mcD_A\mcD^B V^{+(3)}_{B}
-\frac{1}{24}(\Delta_s+ 5)( \Delta_s-1) V^{+(3)}_{A}
-\frac{5}{36}\kappa^{(0)} (\Delta_s+ 3)\mcD_A\mcD^B V^{+(2)}_{B}
\\
&
+ \frac{1}{18}\kappa^{(0)}\mcD_A\mcD^BV^{+(2)}_{B}
+\frac{5}{72}\kappa^{(0)}(\Delta_s+ 3)( \Delta_s-1) V^{+(2)}_{A}
-\frac{1}{36}\kappa^{(0)}(\Delta_s-1)  V^{+(2)}_{A}
\\
&
+ 4mV^{+(2)}_A r^3
+\frac{1}{2} (\kappa^{(0)})^2(\Delta_s+ 3)L_A
-2\kappa^{(0)}c^{(2,0)}_A
\Big]r^3 
+ \mathfrak{O}(r^4)
\,.
\end{align*}
Finally we find
\begin{align*}
(r\partial_{r}-3 + 3\kappa^{(0)} r)\partial_{\tau} V^-_{AB} |_{\scri^-}
=&
  \Big( 3\widehat\Gamma_C{}^0{}_{(A} \partial_{\tau}U_{B)}{}^C
+\partial_{\tau}\check\nabla_{(A}W^+_{B)}
-\kappa ^{(0)}r ^2(\partial_{r}+2\kappa^{(0)}) V^-_{AB} 
\\
&
\hspace{-6em}
+  \frac{1}{2}\Big(\partial_r-\frac{2}{r}+\kappa^{(0)}\Big)\Xi_{ C(A}  V^-_{B)}{}^C
-\frac{3}{4}(\partial_r+\kappa^{(0)})\Xi_{ C(A}   U_{B)}{}^C
-\frac{3}{2r}\mcD^C\Xi_{ C(A}W^+_{B)}
\Big)_{\mathrm{tf}}
\,,
\end{align*}
from which we deduce (it follows from the considerations on $I$, cf.\ Section~\ref{sec_trans_I} below, that  $\partial_{\tau}U_{AB}|_{\scri^-}=O(r)$, alternatively
this can directly be computed from \cite[Equations (2.90)-(2.91)]{ttp_i0}),
\begin{align*}
&\hspace{-2em}
(r\partial_{r}-3 + 3\kappa^{(0)} r)\partial_{\tau} V^-_{A} |_{\scri^-}
\\
=&
\frac{1}{2}  (\Delta_s+ 1)\partial_{\tau}W^+_{A}
+\kappa ^{(0)}(\Delta_s+ 1)L_{ A}  r ^2 
+2\Big((\kappa ^{(0)})^2 (\Delta_s+ 1)L_{ A} -\kappa ^{(0)} c^{(2,0)}_A-4m V^{+(2)}_{A}\Big) r^3 + \mathfrak{O}(r^4)
\\
=&
\frac{1}{2}  (\Delta_s+ 1) (\Delta_s- 1)L_{ A} r
\\
&
 -\frac{1}{2}  (\Delta_s+ 1) \Big( c^{(2,0)}_A+ \mcD_A \mcD^BV^{+(2)}_{B}-  \frac{1}{2}(\Delta_s-1) V^{+(2)}_{A}-4\kappa^{(0)}L_A
\Big)r^2
\\
&
- \frac{1}{2}  (\Delta_s+ 1)\Big[
\frac{1}{12} (\Delta_s+ 5)\mcD_A\mcD^B V^{+(3)}_{B}
-\frac{1}{24}(\Delta_s+ 5)( \Delta_s-1) V^{+(3)}_{A}
-\frac{5}{36}\kappa^{(0)} (\Delta_s+ 3)\mcD_A\mcD^B V^{+(2)}_{B}
\\
&
+ \frac{1}{18}\kappa^{(0)}\mcD_A\mcD^BV^{+(2)}_{B}
+\frac{5}{72}\kappa^{(0)}(\Delta_s+ 3)( \Delta_s-1) V^{+(2)}_{A}
-\frac{1}{36}\kappa^{(0)}(\Delta_s-1)  V^{+(2)}_{A}\Big]
\\
&
+ \Big(\kappa^{(0)} (\Delta_s- 1)c^{(2,0)}_A
-\frac{1}{4} (\kappa^{(0)})^2  (\Delta_s+ 1)(\Delta_s-5)L_A
-  2m (\Delta_s+ 5)V^{+(2)}_A
\Big)r^3 
 + \mathfrak{O}(r^4)
\,.
\end{align*}
The solution is
\begin{align*}
\partial_{\tau} V^-_{A} |_{\scri^-}
=&-\frac{1}{4}  (\Delta_s+ 1) (\Delta_s- 1)L_{ A}r
+\frac{1}{2}  (\Delta_s+ 1)\Big[ \Big( c^{(2,0)}_A+ \mcD_A \mcD^BV^{+(2)}_{B}-  \frac{1}{2}(\Delta_s-1) V^{+(2)}_{A}\Big)
\\
&
-\frac{3}{2}\kappa^{(0)}   \Big(\Delta_s+\frac{5}{3}\Big)L_{ A}
\Big]r^2
+ c_A^{(3,1)}r^3+\mathfrak{O}(r^4)
\,,
\end{align*}
supposing that the following no-logs condition holds
\begin{align*}
&\hspace{-2em}
 \frac{1}{24}  (\Delta_s+ 5) (\Delta_s+ 1)\mcD_A\mcD^B V^{+(3)}_{B}
-\frac{1}{48} (\Delta_s+ 5) (\Delta_s+ 1)( \Delta_s-1) V^{+(3)}_{A}
+ \frac{1}{2}\kappa^{(0)} (\Delta_s+5)c^{(2,0)}_A
\\
=& 
- \frac{1}{2}  (\Delta_s+ 1)\Big[-\frac{5}{36}\kappa^{(0)} (\Delta_s+ 3)\mcD_A\mcD^B V^{+(2)}_{B}
+ \frac{55}{18}\kappa^{(0)}\mcD_A\mcD^BV^{+(2)}_{B}
\\
&
+\frac{5}{72}\kappa^{(0)}(\Delta_s+ 3)( \Delta_s-1) V^{+(2)}_{A}
-\frac{55}{36}\kappa^{(0)}(\Delta_s-1)  V^{+(2)}_{A}\Big]
\\
&
+ (\kappa^{(0)})^2  (\Delta_s- 1)(2\Delta_s-5)L_A
-  2m (\Delta_s+ 5)V^{+(2)}_A
\,.
\end{align*}
Using the no-logs condition  \eq{main_no-logs1}
of the previous order this can be written as
\begin{align*}
&
  (\Delta_s+ 5) (\Delta_s+ 1)\mcD_A\mcD^B V^{+(3)}_{B}
-\frac{1}{2} (\Delta_s+ 5) (\Delta_s+ 1)( \Delta_s-1) V^{+(3)}_{A}
+12\kappa^{(0)} (\Delta_s+5)c^{(2,0)}_A
\\
&
= (\kappa^{(0)})^2  (43\Delta_s^2-72\Delta_ s+ 215)L_A
-  48 m (\Delta_s+ 5)V^{+(2)}_A
\,.
\end{align*}
Divergence and curl, respectively, read
\begin{align*}
(\Delta_s+ 6)\Big( (\Delta_s+ 2)\Delta_s\mcD^A V^{+(3)}_{A}
+24\kappa^{(0)} \mcD^Ac^{(2,0)}_A+ 96m\mcD^AV^{+(2)}_A\Big)
=&2 (\kappa^{(0)})^2 (43\Delta_s^2+14\Delta_s +186) \mcD^AL_A
\,,
\\
 (\Delta_s+ 6) \Big((\Delta_s+ 2) \Delta_s\mcD_{[A} V^{+(3)}_{B]}
-24\kappa^{(0)} \mcD_{[A}c^{(2,0)}_{B]}
-96 m \mcD_{[A}V^{+(2)}_{B]}\Big)
=&-2 (\kappa^{(0)})^2 (43\Delta_s^2+14\Delta_s +186) \mcD_{[A} L_{B]}
\,.
\end{align*}
This can be read as an equation for e.g.\  $c^{(2,0)}_{AB}$, supposing that $L_A$ does not
contain $\ell=2$-spherical harmonics in its decomposition. If it does, the no-logs condition cannot be satisfied.
Note that if $L_A$ does not contain $\ell=2$-spherical harmonics it follows from \eq{main_no-logs1} that $V^+_{AB}$ is not allowed to contain 
$\ell=2$-spherical harmonics in its Hodge decomposition scalars either.
To what extent the no-logs conditions of higher order impose further restrictions on $V^{+(2)}_{AB}$ and $L_A$ is to be analyzed elsewhere.

\begin{lemma}
The no-logs condition which arises from the first-order transverse derivatives requires  that the Hodge decomposition scalars
of $V^+_{AB}$ do not contain $\ell=2$-spherical harmonics  in their harmonic decomposition.
\end{lemma}

\begin{remark}
{\rm
In \cite{ttp_i0} it has been shown that if the radiation field vanishes at all orders at $I^-$ the no-logs conditions are satisfied at each  order at $I^-$ in
an \emph{asymptotically Minkowski-like conformal Gauss gauge at each order} (this is one which fulfills \eq{standard_gauge} below at each order at $I^-$),
from which it follows  that no log-terms are produced in any weakly asymptotically Minkowski-like conformal Gauss gauge). 
Here, even in a simple gauge such as \eq{spec_gauge1}-\eq{spec_gauge3} it is not clear how sufficient conditions look like
as due to e.g.\ \eq{spec_gauge_global_terms} we do not have the polynomial structure of the source terms unless  \eq{the_gauge_cond}
holds, which was crucial in \cite{ttp_i0}.
More general, in a gauge with $\Sigma\ne 0$ an analysis similar to the one in \cite{ttp_i0} seems to be significantly more difficult.

Note that in comparison with a asymptotically Minkowski-like conformal Gauss gauge we do get restrictions on $L_A$, though, in principle, there could be less restrictions
on the expansion coefficients of the radiation field, whence it is not clear whether the set of data which does not produce log terms at $I^-$ is properly
smaller in this gauge, which might be better suitable for an analysis at $I^+$ as the Schwarzschild example suggests
(note that \eq{Schwarzschild_metric_alt} is singular at $\scri^+$).
}
\end{remark}

\subsection{Solution of the transport equations on $I^-$ and time symmetry}
\label{sec_trans_I}

Let us  study  a general  weakly asymptotically Schwarzschild-like conformal Gauss gauge in somewhat more detail.
In the $0$th-order the solutions of the transport equations on $I^-$ are the same as 
for the weakly asymptotically Minkowski-like conformal Gauss gauge computed in \cite{ttp_i0}
as recalled in \eq{Mink_gauge1}-\eq{data_Weyl_I}.
%
%

It is useful to determine the next order, as well. 
First of all note that
\begin{align*}
\Theta =& r(1-\tau^2) + \mathfrak{O}(r^2)
\,,
\\
b_0 =&-2\tau r+ \mathfrak{O}(r^2)
\,,
\quad
b_1 = 2r + \mathfrak{O}(r^2)
\,,
\quad
b_A= 0
\,,
\end{align*}
Connection and frame coefficients, and the Schouten tensor are now obtained from
\eq{transport1}-\eq{transport22}
for initial data at $I^-$ induced by
the limit to $I^-$ of the corresponding fields on $\scri^-$
(which we have determined in Section~\ref{sec_field_on_scri}),
\begin{align*}
\partial_r \widehat L_{10} |_I
=&4m\tau -( \kappa^{(0)}-\theta^{+(0)}-4m)
\,,
\\
\partial_r\widehat L_{11} |_I
=& 2m(1-\tau^2)
\,,
\\
\partial_r\widehat L_{A0} |_I=&\partial_r \widehat L_{A1} |_I=\partial_r\widehat L_{1A} |_I
=0
\,,
\\
\partial_r\widehat L_{AB} |_I
=& -m(1-\tau^2 )\eta_{AB} + \frac{1}{4}(\theta^{-(2)}  -3\theta^{+(0)}-4f_1^{(1)})\eta_{AB}
\,,
\end{align*}
\begin{align*}
\partial_r\widehat\Gamma_{1}{}^0{}_1 |_I
 =&
4m \Big(\tau -\frac{1}{3}\tau^3\Big)    - \Big(\kappa^{(0)} + f_1^{(1)}-\frac{8}{3}m\Big)
\,,
\\
\partial_r\widehat\Gamma_{1}{}^1{}_1 |_I
 =&
-\frac{1}{3}m(1- \tau^4)  + \Big( f_1^{(1)}+\theta^{+(0)}+\frac{4}{3}m\Big)(1+\tau) + f_1^{(1)}
\,,
\\
\partial_r\widehat\Gamma_{1}{}^0{}_A |_I=&\partial_r\widehat\Gamma_{1}{}^1{}_A |_I=\partial_r\widehat\Gamma_{A}{}^0{}_1|_I
=\partial_r\widehat\Gamma_{A}{}^1{}_1 |_I
 =
0
\,,
\\
\partial_r\widehat\Gamma_{A}{}^0{}_B |_I
 =&
 -2m\Big(\tau-\frac{1}{3}\tau^3\Big)\eta_{AB} + \frac{1}{4}(\theta^{-(2)}  -3\theta^{+(0)}-4f_1^{(1)})\tau \eta_{AB}
-\Big(\theta^{+(0)}+f_1^{(1)}+ \frac{4}{3}m\Big) \eta_{AB}  
\,,
\\
\partial_r \widehat\Gamma_{A}{}^1{}_B |_I
 =&
-2\Big[ m\Big(-\frac{5}{6}+ \tau^2-\frac{1}{6}\tau^4\Big)+ \frac{1}{8}(\theta^{-(2)}  -3\theta^{+(0)}-4f_1^{(1)})(1-\tau^2) 
\\
&
+\Big(\theta^{+(0)}+f_1^{(1)}+ \frac{4}{3}m\Big) (1+\tau)+ \frac{1}{4}\Big( \theta^{-(2)} -\theta^{(0)+} -4f_1^{(1)}\Big)\Big]\eta_{AB}
\,,
\\
(\partial_r\widehat\Gamma_{1}{}^A{}_B)_{\mathrm{tf}} |_I
 =&
0
\,,
\\
\partial_r\widehat\Gamma_{A}{}^B{}_C |_I
 =&
 \mathring \Gamma_A{}^B{}_C\Big[ m\Big(-\frac{5}{6}+ \tau^2-\frac{1}{6}\tau^4\Big)+ \frac{1}{8}(\theta^{-(2)}  -3\theta^{+(0)}-4f_1^{(1)})(1-\tau^2) 
\\
&
+\Big(\theta^{+(0)}+f_1^{(1)}+ \frac{4}{3}m\Big) (1+\tau) -\frac{1}{2}\theta^{+(0)} \Big]
\,,
\end{align*}
\begin{align*}
\partial_re^{\tau}{}_1|_I=& \frac{1}{3}m\Big(\tau+4\tau^3 -\tau^5\Big) 
 - \frac{1}{2}\Big( \theta^{+(0)}+ \kappa^{(0)} +2 f_1^{(1)}-\frac{4}{3}m\Big)\tau^2
 - \Big(2 f_1^{(1)}+\theta^{+(0)}+\frac{4}{3}m\Big)\tau
\\
&
 +\frac{1}{2}\Big( \kappa^{(0)} - \theta^{+(0)}-2 f_1^{(1)}-\frac{4}{3}m \Big)
\,,
\\
\partial_re^{r}{}_1|_I=&
1
\,,
\\
\partial_re^{\mathring A}{}_1|_I=& \partial_re^{\tau}{}_A|_I=\partial_re^r{}_A|_I=
0
\,,
\\
\partial_re^{\mathring A}{}_A|_I=& 
\mathring  e^{\mathring A}{}_A\Big[ m\Big(-\frac{5}{6}+ \tau^2-\frac{1}{6}\tau^4\Big)+ \frac{1}{8}(\theta^{-(2)}  -3\theta^{+(0)}-4f_1^{(1)})(1-\tau^2) 
\\
&
+\Big(\theta^{+(0)}+f_1^{(1)}+ \frac{4}{3}m\Big) (1+\tau) -\frac{1}{2}\theta^{+(0)} \Big]
\,.
\end{align*}
The solution will be manifestly time-symmetric up to and including this order if and only if
\begin{align*}
 \kappa^{(0)}-\theta^{+(0)}=&m
\,,
\\
 \kappa^{(0)} + f_1^{(1)}=& \frac{8}{3}m
\,,
\\
 \kappa^{(0)} - \theta^{+(0)}-2 f_1^{(1)} =& \frac{4}{3}m
\,,
\end{align*}
equivalently
\begin{equation}
\kappa^{(0)}=\frac{4}{3}m\,,\quad \theta^{+(0)}=-\frac{8}{3}m\,, \quad f_1^{(1)}=\frac{4}{3}m
\,,
\end{equation}
in consistency with the Schwarzschild gauge determined above.

\begin{lemma}
{\rm
In any gauge where \eq{the_gauge_cond} holds,  solutions with $m=\mathrm{const.}\ne 0$  cannot  be manifestly time-symmetric.
}
\end{lemma}

\section{Gauge transformations}
\label{sec_gauge}

It has been shown in \cite{ttp_i0}  that the gauge transformation from
any weakly asymptotically Minkowski-like conformal Gauss gauge to any other one is smooth at $I^-$.
Here we aim to analyze the  behavior of the gauge transformation near $I^-$ if \eq{the_gauge_cond} is violated.

Given a (smooth) solution to the GCFE in a conformal Gauss gauge with gauge data as specified by \eq{rel_gauge_data_alt},
it has been shown in \cite[Section~6]{ttp_i0} how to transform into any other  conformal Gauss gauge
as specified by a different  gauge data set  \eq{rel_gauge_data_alt}.
The analysis there shows that the only obstruction for the gauge transformation to be smooth at $I^-$ arises
from the rescaling of the $r$ coordinate on $\scri^-$, i.e.\ from the change of the parameterization of the
null geodesic generators of $\scri^-$.

Given a solution to the GCFE with gauge data  $\widetilde\nu_{\tau}$, $\widetilde\Theta^{(1)}$ and $\widetilde\kappa$,  one transforms to the prescribed values of $\nu_{\tau}$, $\Theta^{(1)}$ and $\kappa$ simultaneously \cite{ttp_i0}, and this is accomplished
by a transformation of the form
\begin{equation*}
\widetilde r \mapsto r= r(\widetilde r,x^{\mathring A}) 
\,,
\qquad
\widetilde \Theta \mapsto \Theta = \psi(\widetilde r, x^{\mathring A})\widetilde \Theta
\,, 
\quad 
1+\widetilde \tau \mapsto  1+\tau =h(\widetilde r, x^{\mathring A})(1+\widetilde \tau)
\,,
\end{equation*}
where the function $r$ is determined from the ODE (we suppress dependency on the $x^{\mathring A}$'s henceforth, note that there is a sign error
in the corresponding formula in \cite{ttp_i0})
\begin{equation}
\frac{\partial^2 r}{\partial \widetilde r^2}
=\frac{\partial r}{\partial \widetilde r}\big[\widetilde\kappa(\widetilde r)+2\partial_{\widetilde r}\log\psi(\widetilde r)\big]
  - \Big(\frac{\partial r}{\partial \widetilde r}\Big)^2\kappa( r(\widetilde r))
\,,
\label{crucial_trafo_ODE}
\end{equation}
and where
\begin{equation*}
\psi(\widetilde  r) =h(\widetilde  r)\frac{\Theta^{(1)}(r( \widetilde  r))}{ \widetilde \Theta^{(1)}(\widetilde  r)}
\,, \quad 
h( \widetilde  r)=\frac{\partial \widetilde  r}{\partial r} \frac{ \nu^{\tau}(r(\widetilde  r))}{\widetilde  \nu^{\tau}(\widetilde  r)}
\psi(\tilde r)^2
\,.
\end{equation*}
In a weakly asymptotically Schwarzschild-like conformal Gauss gauge \eq{genPgauge1}-\eq{genPgauge8}  we have
\begin{align*}
\psi(\widetilde r) 
=&
\frac{\partial r}{\partial\widetilde r} \frac{ \widetilde \nu^{\tau}(\widetilde r)}{\nu^{\tau}(r(\widetilde r))}\frac{\widetilde  \Theta^{(1)}(\widetilde r)}{\Theta^{(1)}(r(\widetilde r))}
\\
=&
\frac{\partial r}{\partial\widetilde r} \Big(2\widetilde r^2 +(\widetilde \Theta^{(1,2)}- 2\widetilde \nu_{\tau}^{(0)})\widetilde r^3+ \mathfrak{O}(\widetilde r^4)\Big)
\Big(\frac{1}{2r^2}-\frac{1}{4r}(\Theta^{(1,2)}  - 2\nu_{\tau}^{(0)})  + \mathfrak{O}( r^0)\Big) 
\,,
\end{align*}
i.e.\
\begin{align*}
\partial_{\widetilde r}\log\psi(\widetilde r) 
=&
\frac{\partial\widetilde r}{\partial r}\frac{\partial^2 r}{\partial\widetilde r^2}
+ \frac{2}{\widetilde r} +\frac{1}{2}(\widetilde \Theta^{(1,2)}- 2\widetilde \nu_{\tau}^{(0)})+ \mathfrak{O}(\widetilde r)
+\frac{\partial r}{\partial\widetilde r}
\Big(-\frac{2}{r}-\frac{1}{2}(\Theta^{(1,2)}  - 2\nu_{\tau}^{(0)})  +  \mathfrak{O}( r)\Big) 
\,,
\end{align*}
whence \eq{crucial_trafo_ODE} adopts the form
\begin{align*}
\frac{\partial^2 r}{\partial \widetilde r^2}
=&
\Big(\frac{\partial r}{\partial \widetilde r}\Big)^2
\Big(\frac{2}{r}+ \underbrace{\Theta^{(1,2)}  - 2\nu_{\tau}^{(0)}+ \kappa^{(0)}}_{=\theta^{+(0)} -\kappa^{(0)}=\Sigma} +  \mathfrak{O}( r)\Big) 
-\frac{\partial r}{\partial \widetilde r}\Big(\frac{2}{\widetilde r} 
 +\underbrace{\widetilde \Theta^{(1,2)}- 2\widetilde \nu_{\tau}^{(0)}+ \widetilde\kappa^{(0)} }_{=\widetilde\theta^{+(0)} -\widetilde\kappa^{(0)}=\widetilde \Sigma}+ \mathfrak{O}(\widetilde r)\Big)
\,.
\end{align*}
Equivalently,
\begin{align*}
r\frac{\partial^2 r}{\partial \widetilde r^2}
=
  \Big(\frac{\partial r}{\partial \widetilde r}\Big)^2\Big(2 +\Sigma r+\mathfrak{O}( r^2)\Big)
- r\frac{\partial r}{\partial \widetilde r}\Big(\frac{2}{\widetilde r} +\widetilde \Sigma
+\mathfrak{O}(\widetilde r)\Big)
\,.
\end{align*}
As in \cite{ttp_i0} we set
$u:=\partial_{\widetilde r}\log(\widetilde r/r)$ and $v:=r/\widetilde r$. This way we are led to the system
\begin{align*}
\partial_{\widetilde r} u 
=&
\frac{1}{\widetilde r}(\widetilde \Sigma- \Sigma v)
-u^2 -\widetilde \Sigma  u   +  \Sigma uv (2- \widetilde r u) 
+  v^2(1-u\widetilde r)^2\mathfrak{O}((v\tilde r)^0)
+(1-u\widetilde r)\mathfrak{O}(\widetilde r^0)
\,,
\\
\partial_{\widetilde r} v=&-uv
\,.
\end{align*}
The gauge condition $g_{\mathring A\mathring B}|_{I^-}=s_{\mathring A\mathring B}$ requires \cite{ttp_i0}  $v|_{I^-}=1$,
and it follows from the first equations that a necessary condition for $u$ and $v$ to be smooth at $I^-$ is $\Sigma=\widetilde \Sigma$.
If $\Sigma=\widetilde \Sigma=0$ the equation is regular and this case was considered in \cite{ttp_i0}.
If $\Sigma=\widetilde \Sigma\ne 0$ if follows e.g.\ from \cite{claudel_newman} (if certain technical conditions are added)
that the initial data  $v|_{I^-}=1$ and $u|_{I^-}$ generate smooth solutions.


To sum it up,
the transformation of the $r$ coordinate is the only potential source of log-terms in the gauge transformation.
The decisive quantity with regard to the appearance of logarithmic terms is 
%
\begin{equation*}
\varSigma:=\theta^{(0)}- \kappa^{(0)}
\,.
\end{equation*}
The gauge transformation is only smooth for gauge data with the same $\varSigma$. In other words, whenever one transforms from 
an asymptotically Schwarzschild-like conformal Gauss gauge to another one
the gauge transformation will be non-smooth at $I^-$ if $\varSigma\ne \widetilde\varSigma$.
However, as the Schwarzschild example shows it is still possible that the fields themselves remain regular in \emph{both} gauges at $I^-$. 
This is related to the singular behavior of the metric at $I^-$.
As the Schwarzschild transformation turns out to be rather complicated between the gauges considered above, it is more
illustrative to demonstrate this using a toy model in Section~\ref{sec_toy} while  in Section~\ref{sec_Mink_gauges} we analyze a similar situation for Minkowski in somewhat more detail.

%

\begin{remark}
\label{rem_second_datum}
{\rm
The second datum, $u|_{I^-}$, corresponds to the freedom to prescribe $\partial_{\widetilde r}^2 r|_{I^-}=:q$.
It follows from the computation in \cite[Section~3.4]{ttp_i0} that under the corresponding transformation
 we have for $\Xi_{\mathring A\mathring B}=\mathfrak{O}(r^2)$, 
\begin{align*}
\Xi_{\mathring A\mathring B}\mapsto   & \frac{\partial  r}{\partial \widetilde r}\Big(-2(\widetilde\kappa -\widetilde \theta^+)\rnabla_{\mathring A}\widetilde r\rnabla_{\mathring B}\widetilde r
+2 \widetilde\xi_{(\mathring A}\rnabla_{\mathring B)}\widetilde r
+\widetilde\Xi_{\mathring A\mathring B}
-2\rnabla_{\mathring A}\rnabla_{\mathring B}\widetilde r
\Big)_{\mathrm{tf}}
\\
=  &
\Big(\widetilde\Xi_{\mathring A\mathring B}^{(2)} 
+(\rnabla_{\mathring A}\rnabla_{\mathring B}q )_{\mathrm{tf}}
\Big)\widetilde r^2+ O(\widetilde r^3)
\,.
\end{align*}
In view of \eq{const_mass_aspect} this can be used to transform to a gauge where $\Xi^{(2)}_{\mathring A\mathring B}=0$,
which we assume throughout.
}
\end{remark}

\subsection{A toy model}
\label{sec_toy}

Consider the flat metric in two different coordinate systems,
\begin{equation*}
 g^{(1)}=-\mathrm{d}t^2 + \frac{2}{r^2}\mathrm{d}t\mathrm{d}r +s_{\mathring A\mathring B}\mathrm{d}x^{\mathring A}\mathrm{d}x^{\mathring B}
\,,
\end{equation*}
and
\begin{equation*}
 g^{(2)}=-\mathrm{d}t^2 + \frac{2}{R^2}e^R\mathrm{d}t\mathrm{d}R +s_{\mathring A\mathring B}\mathrm{d}x^{\mathring A}\mathrm{d}x^{\mathring B}
\,.
\end{equation*}
Here we have
\begin{equation*}
\kappa^{(1)} = -\frac{2}{r}\,, \quad \kappa^{(2)}=-\frac{2}{R} + 1
\,.
\end{equation*}
The coordinate transformation which relates $g^{(1)}$ and $g^{(2)}$ is given by
\begin{align*}
\frac{\mathrm{d}r}{r^2} = \frac{e^R}{R^2}\mathrm{d}R
\quad \Longleftrightarrow \quad
\frac{1}{r} =-\int \frac{e^R}{R^2}\mathrm{d}R=\frac{1}{R} -\log R+ O(1)
\quad \Longleftrightarrow \quad
r=R +R^2 \log R+ O(R^2)
\,,
\end{align*}
which is polyhomogeneous at $R=0$, although both line elements admit an expansion at $r=0$ and $R=0$, respectively, in terms of a power series
(note that we have not considered a conformal factor for this model, and that the asymptotic behavior at $r, R=0$ differs by one order as compared to \eq{genPgauge1}).

\subsection{Comparison of both gauges in the case of the Minkowski spacetime}
\label{sec_Mink_gauges}

%
%
%
%

In its standard conformal representation where spatial infinity is represented by a cylinder the Minkowski line element
adopts the form (cf.\ \cite{F_i0, kroon})
\begin{equation}
 \eta = 
-\mathrm{d}\tau^2 - 2\frac{\tau}{  r}\mathrm{d}\tau\mathrm{d}r + \frac{1-\tau^2}{r^2} \mathrm{d}r^2
 +s_{\mathring A \mathring B}\mathrm{d}x^{ \mathring A}\mathrm{d}x^{\mathring  B} 
\,, \quad \Theta=r(1-\tau^2)
\,.
\label{Mink_cyl}
\end{equation}
The Schouten tensor has the following components
\begin{eqnarray*}
L_{\tau\tau} = \frac{1}{2}
\,,
\quad
L_{\tau r} =\frac{\tau}{2r}
\,,
\quad
L_{\tau  \mathring A} = 0
\,,
\\
L_{rr} = -\frac{1-\tau^2}{2r^2}
\,,
\quad
L_{r  \mathring A} = 0
\,,
\quad
L_{ \mathring A\mathring B} = \frac{1}{2}s_{\mathring A\mathring B}
\,.
\end{eqnarray*}
The non-vanishing Christoffel symbols read
\begin{align*}
\Gamma^{\tau}_{\tau\tau} =  \tau
\,,
\quad
\Gamma^{\tau}_{\tau r} =  \frac{\tau^2}{r}
\,,
\quad
\Gamma^{\tau}_{rr} = -\frac{\tau(1-\tau^2)}{r^2}
\,,
\\
\Gamma^{r}_{\tau\tau} =  -r
\,,
\quad
\Gamma^{r}_{\tau r} =  -\tau
\,,
\quad
\Gamma^{r}_{rr} =  -\frac{1+\tau^2}{r}
\,,
\\
\Gamma^{\theta}_{\phi\phi} =  -\sin\theta\cos\theta
\,,
\quad
\Gamma^{\phi}_{\theta \phi} = \cot\theta
\,.
\end{align*}
Recall the conformal geodesics equations \eq{f_eqn1}-\eq{f_eqn2}.
We will be interested in solutions where the angular components vanish identically,
\begin{equation*}
\dot x^{\mathring A}=0
\,, \quad
f_{\mathring A}=0
\,.
\end{equation*}
 Then \eq{f_eqn1}-\eq{f_eqn2} become
\begin{align*}
\dot x^{\nu}\partial_{\nu} \dot x^{\tau} 
+\Gamma^{\tau}_{\nu\alpha}\dot x^{\nu}\dot x^{\alpha} +2 \dot x^{\tau} \dot x^{\rho}f_{\rho}-f^{\tau}  \dot x^{\lambda} \dot x_{\lambda}&= 0
\,,
\\
\dot x^{\nu}\partial_{\nu} \dot x^{r} 
+\Gamma^{r}_{\nu\alpha}\dot x^{\nu}\dot x^{\alpha} +2 \dot x^{r} \dot x^{\rho}f_{\rho}-f^{r}  \dot x^{\lambda} \dot x_{\lambda}&= 0
\,,
\\
\dot x^{\alpha}\partial_{\alpha}  f_{\tau}
-\Gamma^{\lambda}_{\alpha\tau}  \dot x^{\alpha} f_{\lambda}  -f_{\tau} f_{\lambda}\dot x^{\lambda} 
 +\frac{1}{2}  f_{\mu} f^{\mu}\dot x_{\tau} 
&=  L_{\lambda\tau}\dot x^{\lambda}
\,,
\\
\dot x^{\alpha}\partial_{\alpha}  f_{r}
-\Gamma^{\lambda}_{\alpha r}  \dot x^{\alpha} f_{\lambda}  -f_{r} f_{\lambda}\dot x^{\lambda} 
 +\frac{1}{2}  f_{\mu} f^{\mu}\dot x_{r} 
&=  L_{\lambda r}\dot x^{\lambda}
\,.
\end{align*}
For the time being let us further restrict attention to solutions which satisfy
\begin{equation*}
f_{\tau}=0\,, \quad f_r=\frac{1}{r}
\,.
\end{equation*}
Then the conformal geodesics equations become
\begin{align*}
\dot x^{\tau}\partial_{\tau} \dot x^{\tau} + \dot x^{r}\partial_{r} \dot x^{\tau} 
+\frac{2}{r} \dot x^{\tau} \dot x^{r}
&= 0
\,,
\\
\dot x^{\tau}\partial_{\tau} \dot x^{r} + \dot x^{r}\partial_{r} \dot x^{r} 
&= 0
\,.
\end{align*}
In particular, it is immediate to check that $\dot x^{\tau}=1$ and $\dot x^r=0$ solve this system so that
\begin{equation}
\dot x = \partial_{\tau}
\,, \quad
f_{\tau}= 0
\,, \quad
f_r = \frac{1}{r}
\,,\quad
f_{\mathring A}=0
\,,
\label{stand_cong}
\end{equation}
provides a solution to  the conformal geodesics equations  \eq{f_eqn1}-\eq{f_eqn2}, whence  \eq{Mink_cyl} provides a conformal representation of (a subset of) Minkowski spacetime in conformal Gauss coordinates.
%
%
%
In terms of an asymptotic  initial value problem with data on $\scri^-$ the gauge data which generate this gauge are given by
\begin{align}
g_{\tau\tau}|_{\scri^-} =-1\,, \quad 
\nu_{\tau} = \frac{1}{ r}
\,,\quad  \nu_{\mathring A} = 0
\,,
\quad
\kappa=-\frac{2}{ r} 
\,, \quad 
\theta^-=  0
\,, 
\quad
\Theta^{(1)}=2 r
\,,
\nonumber
\\
 f_{\tau}=0\,, \quad  f_r=\frac{1}{ r}\,, \quad  f_{\mathring A}=0
\,, \quad g_{\mathring A\mathring B}|_{I^-}=s_{\mathring A\mathring B}
\,.
\label{standard_gauge}
\end{align}
Although  the Minkowski spacetime does not have a mass, it still provides a simple model to compare conformal Gaussian coordinates
with $\Sigma=0$ and   a gauge \`a la Friedrich with  $\Sigma\ne 0$.
We want to compare, in the coordinates given by \eq{Mink_cyl}, the congruence \eq{stand_cong} of conformal geodesics 
with one where $\Sigma\ne 0$.

It turns out that, even in this relatively simple setting,  the underlying coordinate transformation is fairly complicated.
To gain an idea what is going on we consider gauge data of the form
\begin{align}
\widetilde g_{\tau\tau}|_{\scri^-} =-1\,, \quad 
\widetilde\nu_{\tau} = \frac{1}{\widetilde r}
\,,\quad  \widetilde\nu_{\mathring A} = 0
\,,
\quad
\widetilde\kappa=-\frac{2}{\widetilde r} + 1
\,, \quad 
\widetilde\theta^-=  0
\,, 
\quad
\widetilde\Theta^{(1)}=2\widetilde r
\,,
\nonumber
\\
\widetilde f_{\tau}=0\,, \quad \widetilde f_r=\frac{1}{\widetilde r}\,, \quad \widetilde f_{\mathring A}=0
\,, \quad g_{\mathring A\mathring B}|_{I^-}=s_{\mathring A\mathring B}
\,,
\end{align}
i.e.\ where $\kappa$ is the only gauge datum which differs from those in \eq{standard_gauge}.

This gauge  yields an  asymptotically Schwarzschild-like conformal Gauss gauge with  $\Sigma=1$.
In this gauge  the  data for the conformal geodesics equations are provided by
\begin{equation}
\widetilde{\dot x}|_{\scri^-} = \partial_{\widetilde\tau}
\,, \quad
\widetilde f_{\tau}|_{\scri^-}= 0
\,, \quad
\widetilde f_r|_{\scri^-} =\frac{1}{\widetilde  r}
\,,\quad
\widetilde f_{\mathring A}|_{\scri^-}=0
\,,
\label{compl_cong}
\end{equation}
In the following we compute how these data look like in the standard coordinates \eq{Mink_cyl}.

For this we proceed as described in Section~\ref{sec_gauge} (in this case there is no angular-dependence), i.e.\ we apply a combination of  coordinate and conformal  transformation of the form
\begin{equation*}
\widetilde r \mapsto r= r(\widetilde r) 
\,,
\qquad
\widetilde \Theta \mapsto \widehat \Theta = \psi(\widetilde r)\widetilde \Theta
\,, 
\quad 
1+\widetilde \tau \mapsto  1+\tau =h(\widetilde r)(1+\widetilde \tau)
\,,
\end{equation*}
where 
\begin{align*}
\frac{\partial^2 r}{\partial \widetilde r^2}=&
-\frac{\partial r}{\partial \widetilde r}\Big(\frac{2}{\widetilde r} + 1\Big)
  +\frac{2}{r}\Big(\frac{\partial r}{\partial \widetilde r}\Big)^2
\,,
\quad  \frac{\partial r}{\partial \widetilde r}\Big|_{I^-}=1
\,,
\\
\psi(\widetilde  r) =&\frac{\partial r}{\partial \widetilde  r} \Big(\frac{\widetilde r}{r}\Big)^2
\,, \\
h( \widetilde  r)=& \frac{\partial r}{\partial \widetilde  r} \Big(\frac{\widetilde r}{r}\Big)^3
\,,
\end{align*}
which yields
\begin{align*}
r=&\Big(\frac{1}{\tilde r}e^{-\tilde r} + a + \mathrm{Ei}(-\tilde r)\Big)^{-1}
\\
=& \tilde r - \tilde r^2 \log \tilde r + (1-\gamma -a)\tilde r^2 + O(\tilde r^3)
\,,\quad \text{for some $a\in\mathbb{R}$}
\,,
\\
\psi(\widetilde  r) =&\frac{\partial r}{\partial \widetilde  r} \Big(\frac{\widetilde r}{r}\Big)^2
=e^{-\tilde r} 
\,, \\
h( \widetilde  r)=& \frac{\partial r}{\partial \widetilde  r} \Big(\frac{\widetilde r}{r}\Big)^3
= \frac{\widetilde r}{r}e^{-\tilde r} 
= 1 + \tilde r\log\tilde r + (\gamma+ a-2)\tilde r + O(\tilde r^2\log\tilde r)
\,,
\end{align*}
where $\mathrm{Ei}(r)$ denotes the exponential integral function, and  where $\gamma$ is the Euler constant.
The precise value of $a$ does not matter for our purposes.
In the new gauge we have the following gauge data, which we  decorate with $\widehat\cdot$,
\begin{align*}
\widehat g_{\tau\tau}=-\Big(\frac{ r}{\widetilde r(r)}\Big)^2 \,, \quad 
\widehat\nu_{\tau} = \frac{1}{ r}
\,,\quad  \widehat\nu_{\mathring A} = 0
\,,
\quad
\widehat\kappa=-\frac{2}{ r} 
\,, \quad \widehat \theta^-=  2r^2e^{3\widetilde r(r)}
\,, 
\quad
\\
\widehat\Theta =2r(1+\tau)-  \frac{ r^2}{\widetilde r(r)}e^{\tilde r(r)}  (1+\tau)^2
\,.
\end{align*}
Next, we apply a coordinate transformation of the form (cf.\ \cite[Section~6]{ttp_i0})
\begin{equation*}
r \mapsto   r + p(r)(1+\tau)
\,,\quad
x^{\mathring A} \mapsto    x^{\mathring A}  + q^{\mathring A}(r)(1+\tau)
\,.
\end{equation*}
From
\begin{align*}
\widehat g_{\tau\tau}|_{\scri^-} =&   g_{\tau\tau}
+2p    \nu_{\tau} 
+2 q^{\mathring A}  \nu_{\mathring A} 
+q^{\mathring A}q^{\mathring B}  g_{\mathring A\mathring B}
\,,
\\
\widehat \nu_{\mathring A} |_{\scri^-} =&   \nu_{\mathring A} 
+  q^{\mathring B}  g_{\mathring A \mathring B}
\,,
\end{align*}
we deduce
\begin{align*}
p=\frac{r}{2}\Big[1-\Big(\frac{ r}{\widetilde r(r)}\Big)^2  \Big]
\,,\quad
 q^{\mathring A}=0
\,.
\end{align*}
%
%
Under this coordinate change the expansion in the transverse direction remains invariant,
\begin{align*}
\widehat \theta^-\mapsto  \widehat\theta^-=  2r^2e^{3\widetilde r(r)}
\,.
\end{align*}
We therefore need to apply the conformal transformation (cf.\ \cite[Section~6]{ttp_i0})
%
\begin{align*}
\widehat \Theta\mapsto \Theta=\Big(1+ \frac{r(\widetilde r)}{2} e^{3\widetilde r}(1+\tau)\Big)\widehat\Theta
=\underbrace{\Big(e^{-\tilde r} + \frac{\widetilde r}{2}e^{\tilde r} (1+\widetilde\tau)\Big)}_{=:\psi}\widetilde \Theta
\,,
\end{align*}
in order to transform $\theta^-$  to zero.

The combination of coordinate and conformal transformation considered so far determines how the initial data for the congruence of conformal 
geodesics need to be transformed. We have
\begin{equation*}
\dot x^{\mu}= \frac{\partial x^{\mu}}{\partial \widetilde x^{\alpha}}\widetilde{\dot x}^{\alpha}
\,,
\quad
\widetilde  f_{\mu}-\frac{\partial}{\partial \widetilde x^{\mu}}\log\psi =  \frac{\partial x^{\alpha}}{\partial \widetilde x^{\mu}}f_{\alpha}
\,,
\end{equation*}
which yields
%
\begin{align*}
\dot x^{\tau}|_{\scri^-}=&\frac{\widetilde r}{r(\tilde r)}e^{-\tilde r} 
= 1 + \tilde r \log \tilde r + O(\tilde r)
\,,
\\
\dot x^{r}|_{\scri^-}=& \frac{r(\tilde r)}{2}\Big(\frac{\widetilde r}{r(\tilde r)}-\frac{ r(\tilde r)}{\widetilde r} \Big)e^{-\tilde r} 
=\widetilde r^2\log\widetilde r + O(\tilde r^2)
\,,
\\
\dot x^{\mathring A}|_{\scri^-}=& 0
\,,
\\
f_{\tau} |_{\scri^-}=&-\frac{1}{2}\Big(\frac{\widetilde r}{ r(\tilde r)} -\frac{ r(\tilde r)}{\widetilde r}  \Big) (1+\widetilde r)e^{\tilde r} - \frac{ r(\tilde r)}{2} e^{3\tilde r} 
=-\widetilde r \log\widetilde r + O(\widetilde r)
\,,
\\
f_{r} |_{\scri^-}
=& \frac{\widetilde r(1+\widetilde r)}{ r(\tilde r)^2}e^{\tilde r}
=\frac{1}{\tilde r}  + 2\log\tilde r + O(1)
\,,
\\
 f_{\mathring A} |_{\scri^-}=&  0
\,.
\end{align*}
In fact we are interested in an expansion in terms of $r$ (rather than $\widetilde r$), subject to the equation,
\begin{align*}
\frac{\partial^2\widetilde  r}{\partial  r^2}=&
-\frac{2}{ r} \frac{\partial\widetilde r}{\partial  r}
 + \Big(\frac{2}{\widetilde r}+ 1\Big)\Big(\frac{\partial \widetilde r}{\partial  r}\Big)^2
\,, 
\quad  \frac{\partial \widetilde r}{\partial  r}\Big|_{I^-}=1
\,.
\end{align*}
From this we deduce that the initial data are of the form
\begin{align*}
\dot x^{\tau}|_{\scri^-}=& 1 +  r \log  r + \text{higher orders}
\,,
\\
\dot x^{r}|_{\scri^-}=&  r^2\log r + \text{higher orders}
\,,
\\
\dot x^{\mathring A}|_{\scri^-}=& 0
\,,
\\
f_{\tau} |_{\scri^-}=&- r \log r + \text{higher orders}
\,,
\\
f_{r} |_{\scri^-}
=& \frac{1}{ r} +\text{higher orders}
\,,
\\
 f_{\mathring A} |_{\scri^-}=&  0
\,.
\end{align*}
To conclude, even in this seemingly relatively simple setting, the  gauge transformation
from a $\Sigma\ne0$-conformal Gauss gauge to one with $\Sigma=0$
 looks  fairly  complicated (at develops log-terms at $I^-$).

\section{Deviation equation}
\label{sec_deviation_equation}

Our aim is to find a somewhat more geometric interpretation of $\Sigma$ and the distinction of conformal Gauss gauges
with $\Sigma=0$ and $\Sigma\ne 0$.
For this recall the deviation equation \eq{conf_dev},
\begin{equation*}
a:=\nabla_{\dot x}\nabla_{\dot x} x'=\mathrm{Riem}(\dot x, x')\dot x - S(\nabla_{x'} f)(\dot x, .,\dot x)
-2 S(f)(\dot x,.,\nabla_{\dot x}x')
\,.
\end{equation*}
%
In our conformal Gaussian coordinates we have $x=(1+\tau, r, x^{\mathring A})$
and the deviation equation adopts the form
%
%
\begin{align*}
a_{\mu}=&-R_{\tau r\tau \mu}  - 2g_{\tau\mu}\nabla_r f_{\tau} - \nabla_r f_{\mu}
-2g_{\tau \mu} f^{\nu}\nabla_{\tau}(x')_{\nu} +2  f_{\mu}\nabla_{\tau}(x')_{\tau}
\,.
\end{align*}
That yields
\begin{align*}
\nu^{\tau} a_{i}|_{\scri^-}=&
-R_{0 10 i}  - 2\eta_{0 i}(\nabla_1 -\nabla_0)f_{0} - (\nabla_1-\nabla_0) f_{i}
-2\nu^{\tau} \eta_{0 i} f^a\nabla_{0}(x')_{a} +2  \nu^{\tau} f_{i}\nabla_{0}(x')_{0}
\\
=&
-R_{0 10 i} 
+ 2\eta_{0 i}(f^a + \widehat \Gamma_1{}^a{}_0)f_a
 -\nu^{\tau}\partial_r f_{i}
+(\widehat \Gamma_1{}^a{}_i- S_1{}^a{}_i(f)+ S_0{}^a{}_i(f))f_a
-2\nu^{\tau} \eta_{0 i} f^a\partial_{\tau}(x')_{a} 
\\
&
+2  \nu^{\tau} f_{i}\partial_{\tau} (x')_{0}
-2\nu^{\tau} \eta_{0 i} f^af_a (x')_{0} 
+2  \nu^{\tau} f_{i}f^a(x')_{a}
\\
=&
-R_{0 10 i} 
 -\nu^{\tau}\partial_r f_{i}
+(\widehat \Gamma_1{}^a{}_i- S_1{}^a{}_i+ S_0{}^a{}_i(f))f_a
\,.
\end{align*}
where we used that $(x')_i|_{\scri^-} = e^{\mu}{}_i x'_{\mu} = e^{\mu}{}_i g_{\mu r}$, whence
\begin{align*}
(x')_0|_{\scri^-} =   \nu_{\tau}
\,,
\quad
(x')_1|_{\scri^-}  =    \nu_{\tau}
\,,
\quad
(x')_A|_{\scri^-}  =  0
\,,
\end{align*}
and $\partial_{\tau} (x')_i |_{\scri^-} = e^{\mu}{}_i \partial_{\tau} g_{\mu r}+\partial_{\tau} e^{\mu}{}_i g_{\mu r}$, whence
\begin{align*}
\partial_{\tau}(x')_0|_{\scri^-} =&   \partial_{\tau} g_{\tau r}
\\
 =&  -\nu_{\tau} f_1  
\,,
\\
\partial_{\tau}(x')_1|_{\scri^-}  =&  \partial_{\tau} g_{\tau r}+\nu^{\tau} \partial_{\tau} g_{r r}+\nu_{\tau}\partial_{\tau} e^{\tau}{}_1 
\\
 =& \nu_{\tau}\widehat\Gamma_{1}{}^1{}_{0} 
\,,
\\
\partial_{\tau}(x')_A|_{\scri^-}  =&    e^{\mathring A}{}_A \partial_{\tau} g_{ r\mathring A}-\nu^{\tau}\nu_A\partial_{\tau} g_{ rr}+\nu_{\tau}\partial_{\tau} e^{\tau}{}_A 
\\
  =& 
  \nu_{\tau}\widehat\Gamma_1{}^0{}_A
\,.
\end{align*}
Here we used that
$\partial_{\tau}g_{\mu r}= -g_{\mu\nu}g_{r\sigma}\partial_{\tau} g^{\nu\sigma}=
-\eta^{ij}g_{\mu\nu}g_{r\sigma}\partial_{\tau}(e^{\nu}{}_ie^{\sigma}{}_j )$, whence
\begin{align*}
\partial_{\tau}g_{\tau r}|_{\scri^-}
=&\nu_{\tau} \partial_{\tau}e^{\tau}{}_1 
 -(\nu_{\tau})^2\partial_{\tau}e^{r}{}_1 
-\nu_{\tau}\nu_{\mathring A}\partial_{\tau}e^{\mathring A}{}_1
\\
 =&-\nu_{\tau} f_1 
\,,
\\
\partial_{\tau}g_{r r}|_{\scri^-} =& -2(\nu_{\tau})^2\partial_{\tau} e^{\tau}{}_1 
\\
=& 2(\nu_{\tau})^2(f_1 +\widehat\Gamma_{1}{}^1{}_{0} )
\,,
\\
\partial_{\tau}g_{\mathring A r}|_{\scri^-} =& -2\nu_{\tau}\nu_{\mathring A  }\partial_{\tau}e^{\tau}{}_1
-\nu_{\tau}g_{\mathring A  \mathring B}\partial_{\tau}e^{\mathring B}{}_1
-\sigma^A{}_{\mathring A}\nu_{\tau}\partial_{\tau}e^{\tau}{}_A 
\\
=&
2\nu_{\tau}\nu_{\mathring A  }( f_1  +\widehat\Gamma_{1}{}^1{}_{0})
+\sigma^A{}_{\mathring A}\nu_{\tau}(f_A +\widehat\Gamma_{A}{}^1{}_{0}+ \widehat\Gamma_1{}^0{}_A )
\,.
\end{align*}
%
%
We thus obtain for the components of $a$,
\begin{align*}
\nu^{\tau} a_{0}|_{\scri^-}
=&
(f_1 + \widehat \Gamma_1{}^1{}_0)f_1+(f^A + \widehat \Gamma_1{}^A{}_0)f_A
\,,
\\
\nu^{\tau} a_{1}|_{\scri^-}
=&
L_{11}-L_{00}
 -\nu^{\tau}\partial_r f_{1}
+(\widehat \Gamma_1{}^A{}_1+ f^A)f_A
\\
=&
\partial_{\tau} f_1 + \widehat L_{11}-f_1f_1
\\
=&
\widehat L_{10}  + \widehat L_{11}  - (f_1 + \widehat \Gamma_1{}^1{}_0)f_1- f_A\widehat\Gamma_{1}{}^A{}_{0} 
\,,
\\
\nu^{\tau} a_{A}|_{\scri^-}
=&
L_{1A} -\nu^{\tau}\partial_r f_{A}
+(\widehat \Gamma_1{}^1{}_A- f_A)f_1
\\
=&
\widehat L_{1A}  +\partial_{\tau}f_A -f_1f_A
\\
=&
\widehat L_{A0}+ \widehat L_{1A}    -  f_1(f_ A+ \widehat\Gamma_{A}{}^1{}_{0})-f_B\widehat\Gamma_{A}{}^B{}_{0}
\,.
\end{align*}
%
In a weakly asymptotically Schwarzschild-like conformal Gauss gauge that yields the following expansions at $I^-$,
\begin{align*}
\nu^{\tau} a_{0}|_{\scri^-}
=&
1+(f_1^{(1)}-\kappa^{(0)})r + \mathfrak{O}(r^2)
\,,
\\
\nu^{\tau} a_{1}|_{\scri^-}
=&
  - 1- (f_1^{(1)}-\theta^{+(0)})r + \mathfrak{O}(r^2)
\,,
\\
\nu^{\tau} a_{A}|_{\scri^-}
=& \mathfrak{O}(r^2)
\,,
\end{align*}
and
\begin{equation*}
a_ia^i|_{\scri^-}=\frac{2}{r}  (\kappa^{(0)}-\theta^{+(0)})+ \mathfrak{O}(1)
\,.
\end{equation*}
A   weakly asymptotically Minkowski-like conformal Gauss gauge is distinguished in the class of weakly asymptotically Schwarzschild-like conformal Gauss gauges
by the property that $a_ia^i|_{\scri^-}$ remains finite at $I^-$. In particular this is the distinctive feature compared to Friedrich's Schwarzschild gauge.

\section*{Acknowledgments}
The author wishes to thank  Helmut Friedrich for useful discussions and comments. Furthermore, the author is  thankful to the Max Planck Institute for Gravitational Physics in Golm, Germany, for hospitality, where part
of the work on this paper has been done.
Financial support by the Austrian Science Fund (FWF) P 28495-N27 is gratefully acknowledged, as well.

\end{document}